# Structure and temporal evolution of transportation literature

Milad Haghani[1*], Michiel C. J. Bliemer[2]

[1]School of Civil and Environmental Engineering, The University of New South Wales, Sydney, Australia
[2]Institute of Transport and Logistics Studies, The University of Sydney, Australia

[*]corresponding author: milad.haghani@unsw.edu.au

## Abstract

Fifty years of evolution of the transportation field is revisited at a macro scale using scientometric analysis of all publications in all 39 journals indexed in the category of 'Transportation' by the Web of Science. The size of the literature, as published by these journals, is estimated to have reached 50,000 documents. The structural composition of this literature is analysed through the lenses of the co-occurrence of key terms, as well as the similarity of article referencing at the level of journals and authors. At the highest level of aggregation, four major divisions of the literature are differentiated through these analyses, namely (i) network analysis and traffic flow, (ii) economics of transportation and logistics, (iii) travel behaviour, and (iv) road safety. Influential and emerging authors of each division are identified. Temporal trends in transportation research are also investigated via document co-citation analysis. This analysis identifies various major streams of transportation research while determining their approximate time of emergence and duration of activity. It documents topics that have been most trendy at any period of time during the last fifty years. Three clusters associated with the travel behaviour division (collectively embodying topics of "land-use", "active transportation", "residential self-selection", "traveller experience/satisfaction", "social exclusion" and "transport/spatial equity"), one cluster of "statistical modelling of road accidents", and a cluster of network modelling linked predominantly to the notion of "macroscopic fundamental diagram" demonstrate characteristics of being *current hot topics* of the field. Three smaller clusters linked predominantly to "electric mobility" and "autonomous/automated vehicles" show characteristics of being *emerging hot topics*. A cluster labelled "shared mobility" is the *youngest emerging* cluster. Influential articles within each cluster of references are identified. Additional outcomes are the determination the *influential outsiders* of the transportation field, articles published by non-transportation journals that have proven instrumental in the development of the transportation literature. These findings provide new and holistic insight into the composition of the transportation field and document its most fundamental studies, while providing objective historical perspectives. Such documentation could also assist subsequent conventional review articles. Analyses on patterns of authorships and co-authorships are expected to facilitate further collaborations among transportation researchers.

## 1. Introduction

### 1.1. Background

Scholarly research in the field of transportation has been documented for more than half a century through its scientific publications. While many articles of review nature have attempted to document the progress made in various sectors of this research field and to synthesise their findings, the field has never been analysed in its entirety. With the significant methodological developments in research synthesis and scientometric methods, such as the ground-breaking document co-citation methodology of Chen (2004), objective analysis of the literature at such a large scale has now been made possible. Harnessing these recent methodological capabilities, here we aim to document fifty years of scientific developments in the field of transportation. Our focus is on the following key questions:

i. What is the estimated size of the transportation literature and how best can the underlying scientometric data of that literature be captured?

ii. What portion of transportation literature is composed of research synthesis (i.e. review) articles?

iii. How can various streams/divisions of this literature be determined on an objective basis and at various levels of aggregations (in other words, what are the major divisions of this field and what are the main research streams within each division)?

iv. What are the temporal relations between various streams of transportation literature and how can their temporal evolution be documented?

v. What are the most influential entities of transportation literature in various divisions of the field?

We investigate the abovementioned main questions by retrieving and analysing all publications in all journals indexed by the Web of Science (WoS) in the category of 'Transportation'. In addition to the main questions laid out above, our analyses also provide historical perspectives on the early stages of the development of the transportation field, identify so-called *hot topics* (i.e., streams of research that have consistently been most active during the recent years), and *influential outsiders* (articles that have not been published by 'transportation' journals per se, but have been found highly instrumental by articles of 'transportation' journals and have played a noticeable role in development of the transportation knowledge).

This scientometric review provides a comprehensive set of analysis of the scholarly transportation literature, ranging from analyses at the level of journals, terms and keywords, authors and organisations and individual articles. It is also the largest scale scientometric analysis ever undertaken in the field of transportation and arguably one of the largest scale reviews of scientometric nature in general. It is expected that the outcomes provide new insight into the structure and composition of the transportation literature based on objective measures, as well as its temporal patterns of development and its most influential articles. This could also assist subsequent conventional review articles and facilitate author collaborations.

Our analysis is meant to document the history of transport research, i.e. it is a meta-archive that describes the transport research of the past fifty years. One of the key objectives was to provide an overview of past fifty years of research, which establishes a baseline for future studies. We believe that now is the right moment to do such a scientometric analysis because the transportation literature is growing very rapidly (currently exponentially) and it will become increasingly difficult to do such an analysis in the future. Subsequent scientometric studies, to be conducted perhaps once every decade, would then only need to focus on trends in the past ten years. This way, a complete meta-archive of the transport research literature can be established (Najmi et al., 2017).

### 1.2. Reading guidance

Given the enormity of the size of transportation literature and the multitude of analytical angels that were required to be explored in order to provide a comprehensive overview of the field, the paper has to cover a great amount of grounds. Outcomes of Section 4 may be particularly useful to authors for determining right journals for publishing their transportation articles as well as to journal editors and authors for gaining an overview of the trends in journal metrics over the past decades. Readers who wish to obtain an objective understanding of the overall structure of the field at an aggregate level may refer to Section 5. Journal editors may also find outcomes of Section 6 (on emerging and established influential authors of various divisions) useful for identifying potential reviewers for transportation papers. Early-career transportation researchers may also use outcomes of this section for forming and/or expanding their network of collaborations across various divisions of the field. Readers who are interested in obtaining a more in-depth temporal understanding about the development of the field and its influential references may refer to section 7. This sections also provides interesting insights into the recent trends of the field and its future directions. This section could also particularly be useful in informing future conventional reviews or enriching literature review sections of original research articles on various transportation topics.

## 2. Methods and data acquisition strategy

The main search query string was designed to generate all articles of the journals indexed by the WoS in the 'Transportation' category. The official names of the target journals were obtained from the WoS and were separated from one another using the Boolean operator OR while each title placed within quotation marks (see Appendix A). This query string was entered into the Advanced Search section of the WoS by equating it to "SO" which is the WoS search engine code (i.e., Field Tag) for searching the journal/source titles (or more precisely, "Publication Name" as worded by the WoS). No time span or any other constraint was added to the search strategy. This query string search returned N=49,543 items (conducted in July 2020) and can be used as a standard query to generate transportation articles and keep track of future developments of this literature.

A secondary search query was devised based upon the primary query in order to generate a subset of articles of 'transportation' journals that are of research synthesis (review) nature. In doing so a fairly inclusive list of terms (or term combinations) that often characterise such articles were devised and were combined with one another via the Boolean operator OR and being equated to "TI" which is the WoS Field Tag for searching the Titles of articles. A separate list of terms or term combinations that authors often use in their keyword list to characterise a review article were combined in a similar manner while being equated to "AK", Field Tag for Author Keywords (see Appendix B). This query string itself was combined with the previous query string that searches for Title of articles through the Boolean operator OR. This combined string itself was combined, using Boolean operator AND, with the string of journals' Titles in order to generate the subset of review articles published by 'Transportation' journals. This search returned a total of N=829 items.

Citation and full bibliometric information of articles of both datasets were retrieved from the WoS and stored in the form of text files. This includes the text of their abstracts and keywords and lists of references as well as their list of authors and their affiliation details, year, source title, volume, issue, page and citation count. The main dataset is accessible in the Online Supplementary Material of this article. Additional raw data of citations and Impact Factors of these individual journals were also retrieved separately from the WoS Journal Citation Report.

Note that our search strategy focuses exclusively on journals indexed in the category of 'Transportation' as this is the category where the vast majority of conventional specialty journals that publish transportation research are listed. As such, the category of 'Transportation Science & Technology' which consists of many journals that do not exclusively publish transportation research, (e.g., *Computer-Aided Civil and Infrastructure Engineering* or *IEEE Vehicular Technology Magazine*) was not considered. Fortunately, almost all relevant transportation journals (those that are regarded conventional journals of transportation research) that are indexed in this second category are also indexed in the 'Transportation' category, and hence, already included in our dataset. The only exception is *Transportation Research Record* (TRR) that is only indexed in the categories of 'Transportation Science & Technology' and 'Engineering, Civil'. After consideration of various factors, we decided not to add this journal to the list for the following reasons:

- A substantial portion of publications of this journal are related to soil mechanics and pavement research, which is not the focus of this review;
- Its articles do not index any keywords (which is a major element of the analyses of this study) and as such do not contribute to the keyword co-occurrence analysis;
- The exceptionally large number of items publish by this journal (estimated to be 15,755 items) would dominate those of every other journal and hence, the journal-level analysis.

Based on these considerations and given the magnitude of the literature published by TRR, we believe that the content of this journal could warrant its own scientometric analysis. Therefore, in the Online Supplementary Material of this article, we have provided some limited and raw analyses of the content published by this journal to which interested readers can refer.

The analyses on the frequency and co-occurrence of key terms, the relationship between journals and authors, and the analyses on networks of collaboration all follow the methodology of van Eck and Waltman (2014), *visualisation of similarities* (VOS) as implemented in the scientometric software VOSviewer (Van Eck and Waltman, 2010), while the co-citation analyses at the level of individual articles follow the methodology of Chen (2004) as implemented in the scientometric software CiteSpace (Chen, 2006).

## 3. Historical perspectives and general statistics of the transportation literature

The size of the peer-reviewed transportation literature as defined earlier is estimated to have reached almost 50,000 items, with the earliest articles on the record dating back to 1967. Nearly 54,000 authors from more than 13,600 institutes and organisations have contributed to this literature. Nearly 1.7% of these articles have identified themselves as a form of research synthesis article. For the studies published during the last full year on the record, 2019, the fraction of review articles has been 2.1% (with 97 review articles out of 4,630 transportation publications in 2019). Figure 1 visualises the number of published articles by "Transportation" journals, as well as the number of published review articles, during each year since 1967. Of the total review articles, nearly 19% are estimated to be meta-analyses and only nearly 2% scientometric/bibliometric reviews (Figure 1(b)). All bibliometric reviews are 2015 or newer items. The year 2007 marks the point after which the number of published articles of "Transportation" journals exceeded 1,000 per year, and the beginning of an exponential growth in the size of this literature.

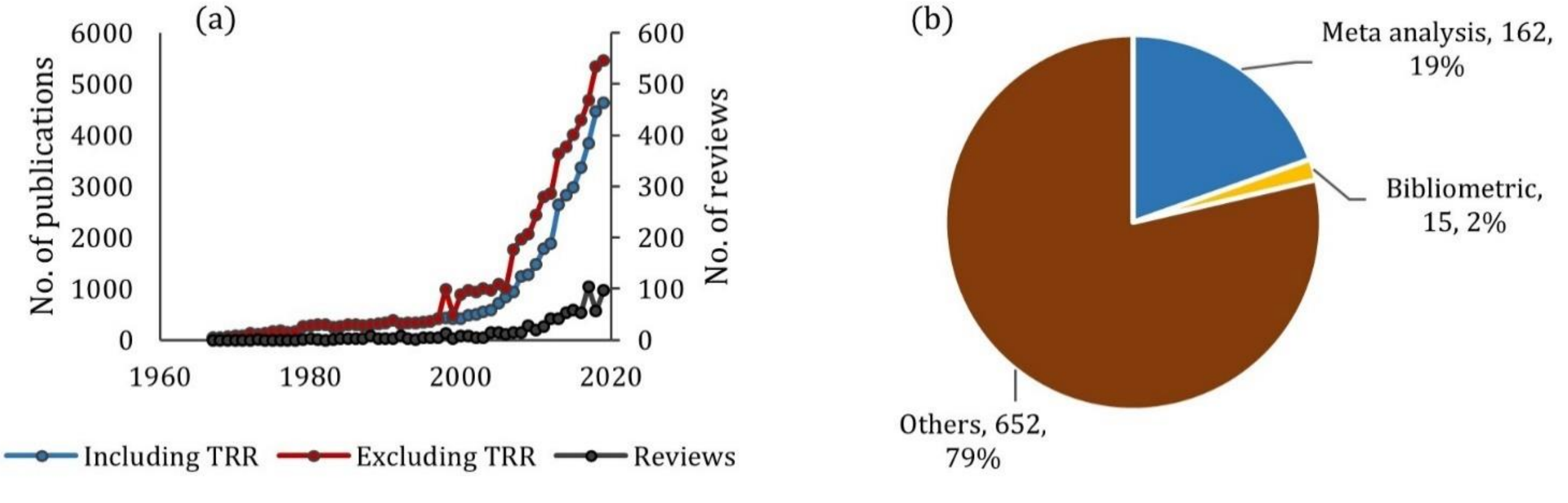

**Figure 1** The number of articles and review articles published by journals indexed as "Transportation" since 1967. Note the scale of the two vertical axes are ten times different.

Earliest articles of this field have predominantly been published by *Journal of Transport Economics and Policy* that could be regarded as the oldest source of transportation articles. In fact, during 1967 and 1968, all publications on the record were published exclusively by this journal. The first article in the first issue of this journal, written by Plowden (1967) is titled "Transportation studies examined". Since 1969, articles also started to emerge in *Journal of Safety Research*, as a then-newly launched journal which continued to become one of the major outlets of the road safety sector of the transportation literature. The first research article in the first issue of this journal (that included 20 articles) is an article by Forbes et al. (1969) entitled "*Low contrast and* *standard visual acuity under mesopic and photopic illumination*". During the 1960s these two journals remained exclusively the only sources of publications in this field, until 1970 that marked the launch of *Transportation Journal*, as the third oldest journal established in this field. The first article of this journal is one authored by Davis (1970) with the title "Modifications in identifying characteristics of several federal transportation activities". The youngest indexed journals in this category are *Journal of Transport and Health*, *Travel Behaviour and Society* and *Analytic Methods in Accident Research*.

Review articles started to emerge in the transportation literature since the 1970s. The first review on the record is a *Journal of Safety Research* article authored by Perrine (1973a) entitled "Alcohol influences on driving-related behavior: A critical review of laboratory studies of neurophysiological, neuromuscular, and sensory activity". It took six more years for the second and third review article of transportation literature to emerge and they were articles written by Train (1979) ("Consumers' responses to fuel-efficient vehicles: a critical review of econometric studies" in *Transportation*) and Williams (1980) ("The National Transportation Policy Study Commission and Its Final Report: A Review" in *Transportation Journal*). By 1985, the size of the subset of review studies in transportation had reached 10 items. With 23 articles of review type (including meta analyses) to his name, Rune Elvik has been the single author who has contributed the most to this subset.

## 4. Journal relationships and metrics in transportation literature

According to our data, an excess of 4,000 articles have been added to the transportation literature every year since 2018 by 'Transportation' journals. The dissemination of transportation studies across these journals is not even. In fact, nearly a quarter of all articles ever published in this field have been collectively published by three journals: "*Accident Analysis and Prevention*", "*Transportation Research Part A: Policy and Practice*" and "*Transportation Research Part B: Methodological*". Figure 2 shows the rate of accumulation of articles in transportation literature across all 39 journals by presenting the number of articles published by each journal during each year since 1967.

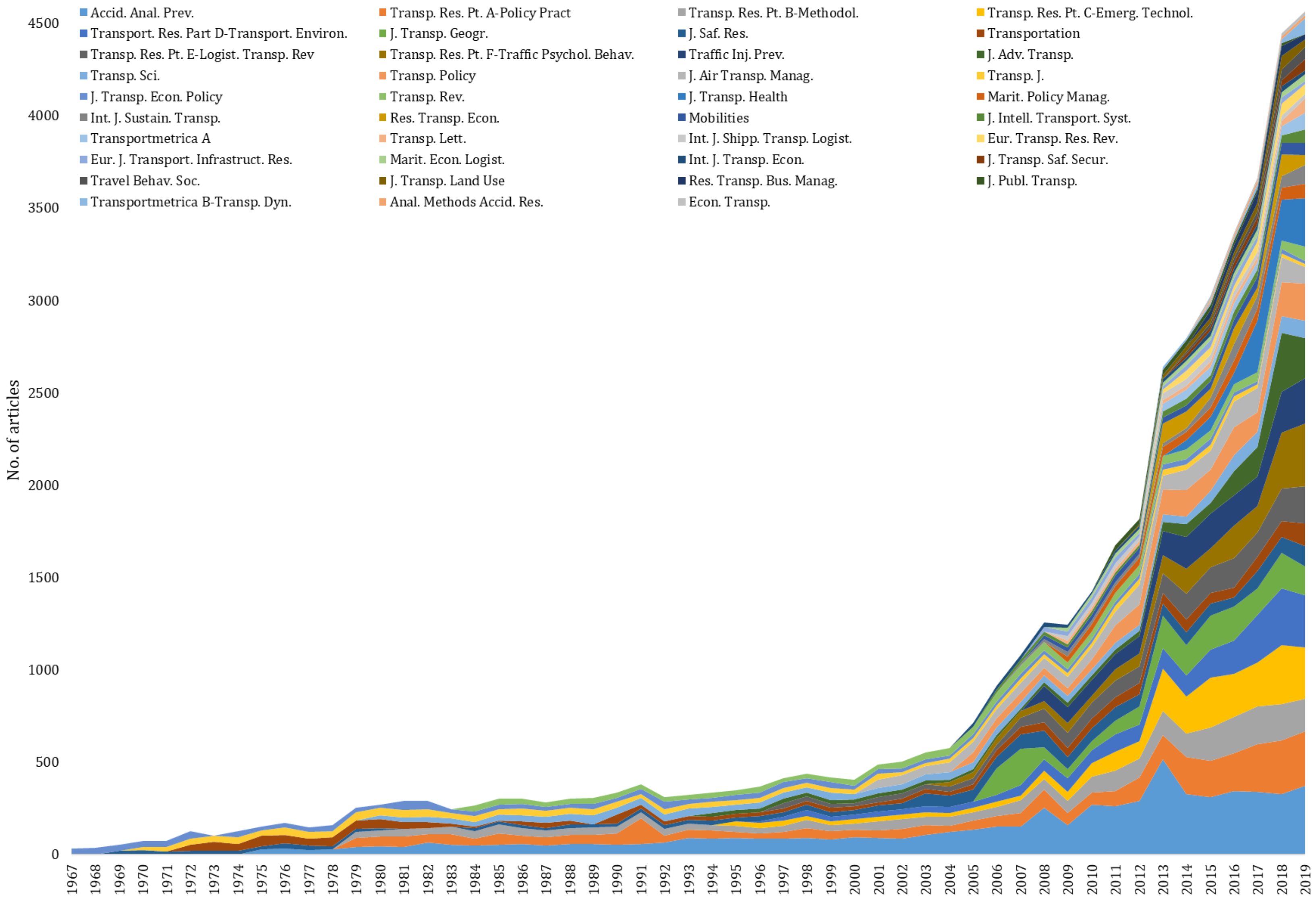

**Figure 2** The number of articles published by each of the 'Transportation' journals over the years from 1967 to 2019.

Figure 3 visualises the data regarding two metrics, as well as their variations over time, that are commonly used as measures of performance or significance of the journals or the impact of their published content in the field. This includes the number of citations to their documents and *journal Impact Factor* (IF), representing the mean citations per paper published by a journal over a two year period (Mingers and Leydesdorff, 2015). This could include or alternatively exclude citations from articles of the same journal itself; constituting IF with and without self cites. In Figure 3, the total number of citations are being represented by continuous red lines and the left vertical axes with display units of thousands. Journals' IF with and without self cites are respectively shown with solid and dashed black lines and are represented by right vertical axes.

According to this data, almost all 'Transportation' journals have been accumulating citations at an increasing rate over the years which could be an indication of a continuously growing field. However, there are exceptions, most notably "*Transportation Journal*", "*International Journal of Transportation Economics*" and "*International Journal of Shipping and Transport Logistics*" whose total number of citation counts have shown significant fluctuations over time. The IF of 'Transportation' journals has been subject to more frequent fluctuations, compared to the number of citations. Despite transient fluctuations, the general trend for the IF of the majority of these journals (28 items) has been increasing. According to data associated with the last full year on the record, 2019, "*Analytic Methods in Accident Research*" (IF=9.179)—as one of the youngest transportation journals—has received the largest average count of citation to its articles published in 2018 and 2017 (an average of 9.179 citations to each item), compared to any other 'Transportation' journal. "*International Journal of Transport Economics*" (IF=0.509), on the other hand, has been the recipient of the smallest normalised counts of citations to its 2018 and 2017 articles among all 'Transportation' journals. The mean and median of IF for 'Transportation' journals associated with the 2019 record are respectively 2.792 and 2.424. For the majority of these journals, the IF with and without self cites have been highly correlated over the years. Exceptions to this pattern are "*Journal of Transport and Health", "Maritime Policy & Management*", and to a lesser degree "*Transportmetrica A: Transport Science*", whose IF with and without self cites have changed at opposite directions during certain periods in time. On the other hand, the IF of certain journals has not only been highly correlated but also consistently similar in quantity over the years. The most noticeable include "*Transportation*", "*Transport Reviews*", "*Transportation Letters*", "*European Journal of Transport and Infrastructure Research*", "*Transportation Science*", "*Transport Policy*", "*Journal of Safety Research*" and "*Journal of Transportation Safety and Security*". Certain other journals such as "*Analytic Methods in Accident Research*", "*Traffic Injury Prevention*", "*Mobilities*", "*Journal of Intelligent Transportation systems"*, "*Transportmetica A*" and "*Journal of Transport and Health*" have consistently maintained a gap between their IF with and without self cites, meaning that a certain portion of the counts of citation to their content comes from studies by the journal itself, which could be a typical pattern for journals that maintain a narrow specialty.

The relation between 'Transportation' journals is further analysed in the following based on the similarity of the references of their published articles; i.e. *journal bibliographic coupling* (Kessler, 1963), a metric that could represent thematic similarities between articles, or at an aggregate level, between publications of journals. The outcomes have been presented in Figure 4. Each node represents a single 'Transportation' journal while the size of the node is proportional to the total number of documents published by the journal. Journals whose publications have more similar references, i.e. those whose publications display a strong bibliographic coupling relation, are generally visualised closer to one another and can form clusters (Haghani, 2021; Haghani et al., 2021; Van Eck and Waltman, 2010). The thickness of the links between individual nodes is also an indicator of the strength of the bibliographic relation between the pair of journals that are connected. A total of four clusters of 'Transportation' journals were identified based on this metric. Cluster #1 (shown blue) with n=14 items is the largest cluster. This is followed by cluster #2 (shown purple) with n=11, cluster #3 (shown red) with n=8 and cluster #4 (shown orange) with n=6. Hybrid maps are also visualised below the main visualisation of the network representing average number of publications years (left) and average number of citations (right) associated with items of each journal.

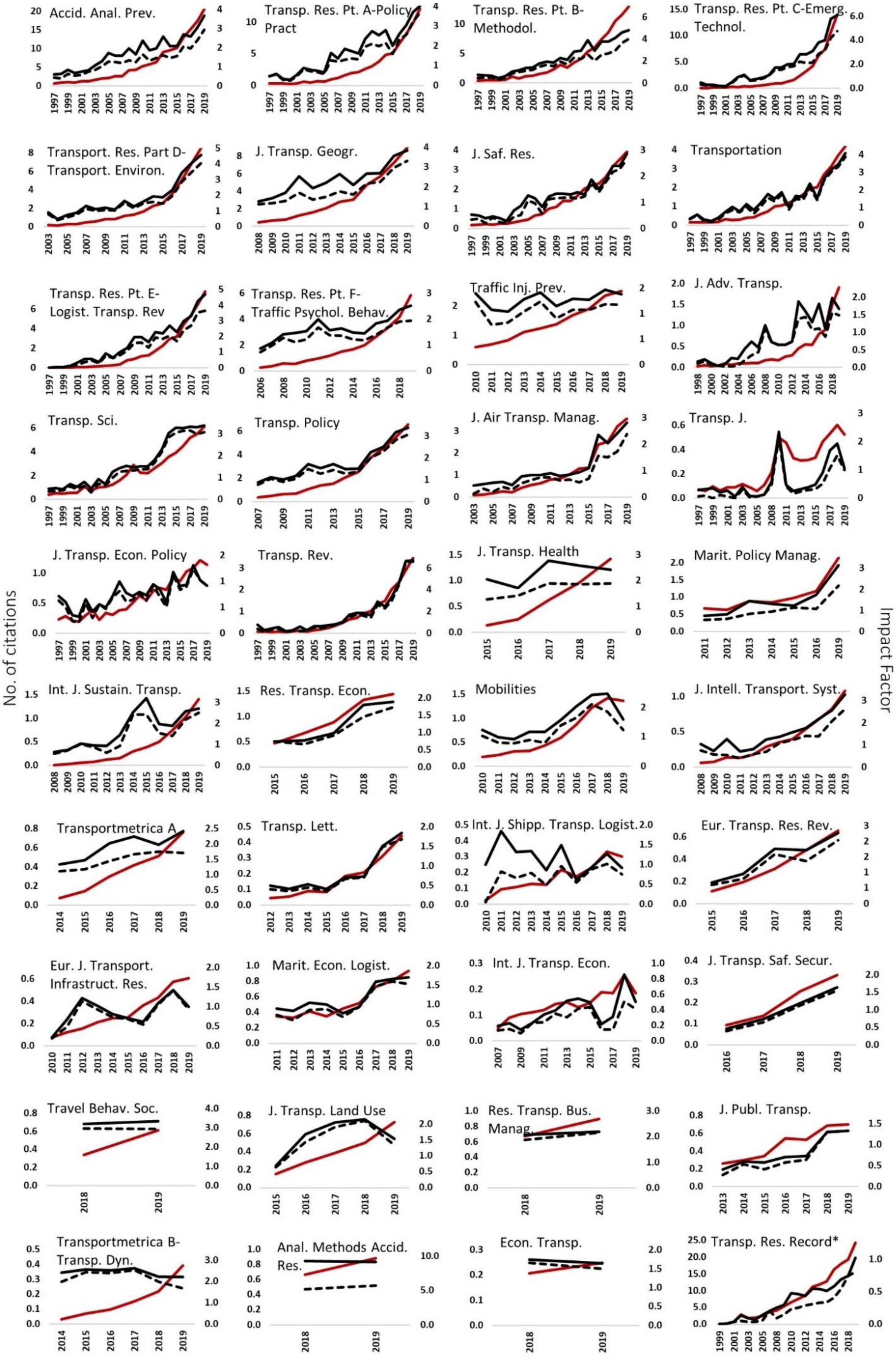

**Figure 3** Citations (red, left axis), journal IF and IF without self-cite (solid and dashed black lines, right axis).

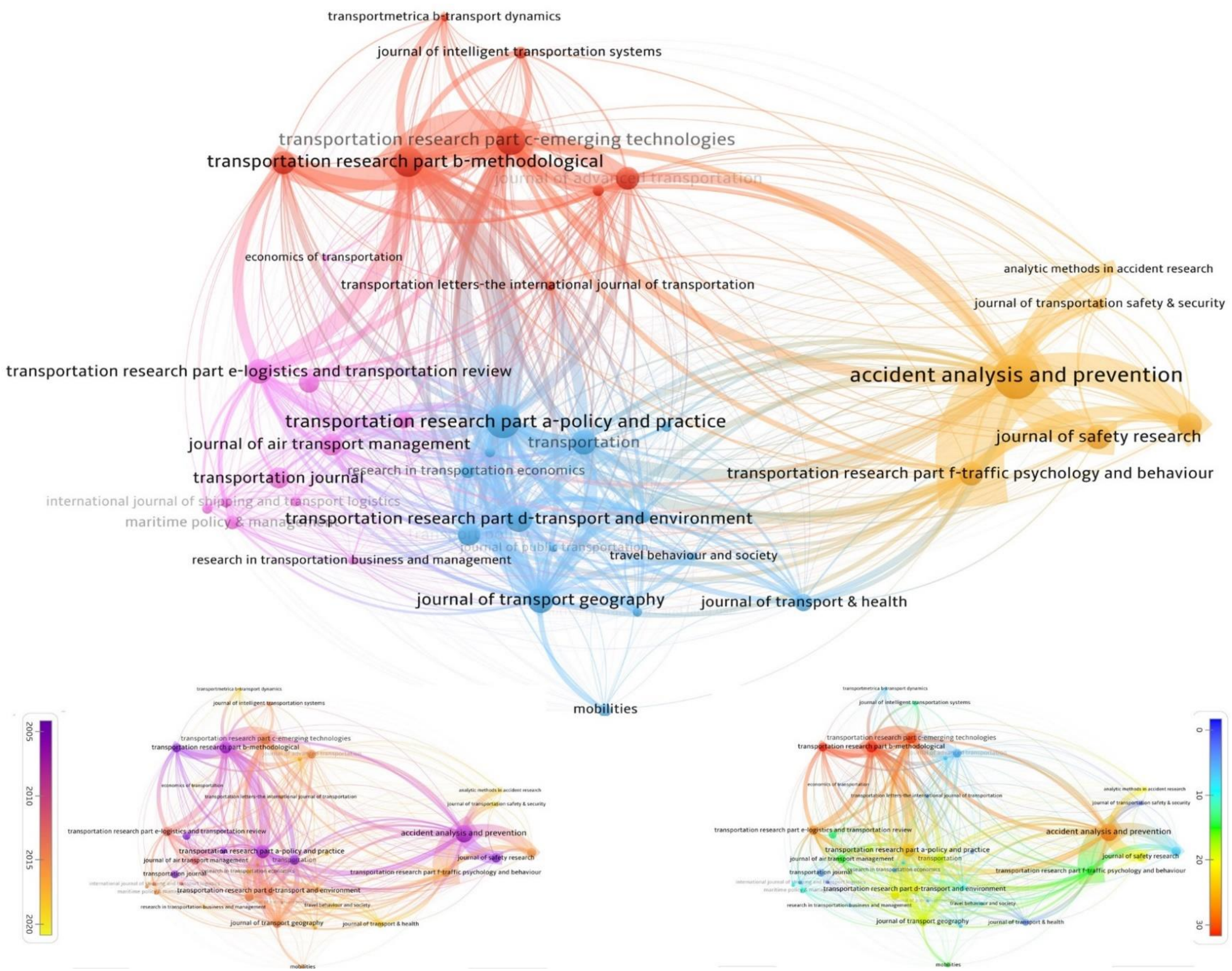

**Figure 4** Bibliographic coupling of 'Transportation' journals (top) where clusters #1, #2, #3 and #4 are respectively shown blue, purple, red and orange; Hybrid map with an overlay of average publication year (bottom left) and hybrid map with an overlay of average citation count (bottom right).

The most dominant items of clusters #1, #2, #3 and #4 are respectively "*Transportation Research Part A: Policy and Practice*" (n=3,715), "*Transportation Research Part E-Logistics and Transportation Review*" (n=1,895), "*Transportation Research Part B-Methodological*" (n=2,884) and "*Accident Analysis and Prevention*" (n=6,410). Journals of cluster #4 (representing those whose publications focus predominantly on road safety) are internally coupled with one another in a visibly strong way while being relatively isolated/distant from the rest of the body of 'Transportation' journals. They form a very clear and internally well-connected cluster whose items are strongly coupled, but the strength of their bibliographic relation with items of other clusters is relatively weak. "*Transportation Research Part A*" on the other hand, being situated near the centre of the map is a single journal that shows distinctly strong bibliographic coupling with journals from all four clusters. This is reflected in the metric of *total link strength* as measured by VOSviewer (Haghani and Bliemer, 2020; Haghani et al., 2020; Van Eck and Waltman, 2010), representing the total strength of the bibliographic relationship of a journal with other journals. "*Transportation Research Part A*" scores the highest at this metric compared to all other 'Transportation' journals. In other words, items of this journal have thematic similarities with those of all other clusters, indicating that this journal has the most diverse range of publication themes compared to all other 'Transportation' journals, one that could be considered to play a *betweenness centrality* role (Leydesdorff, 2007) in transportation research, comparable to a *multidisciplinary* journal in a broader sense. *Transportation* is also another journal of this literature that presents such feature, though to a lesser degree. The highest strength of bibliographic relationship, by far margin, is recorded between "*Accident Analysis and Prevention*" and "*Transportation Research Part F*". Outside cluster #4, however, the strongest relationship exists between "*Transportation Research Part C*" and "*Transportation Research Part B*". Journals with oldest publications (in terms of average year (AY)) are "*Journal of Transport Economics and Policy*" (AY=1991.48), "*Transportation Journal*" (AY=1992.77) and "*Transportation*" (AY=1999.73). In terms of average citation (AC) count, articles of "*Transportation Research Part B*" (AC=37.84) and "*Transportation Science*" (AC=37.38) and "*Transportation Research Part C*" (AC=29.06) score the highest. When normalised for the age of publications, these three journals still stand out the most in addition to "*Analytic Methods in Accident Research*".

## 5. Keyword, title and abstract analysis of transportation articles

One approach to objectively identify segments of transportation literature, perhaps at the most aggregate level, is by identifying patterns and frequencies of the occurrence of terms in the titles, abstracts and keyword lists of its articles. Keyword lists as well as title and abstracts of all N=49,543 'Transportation' articles were analysed for this purpose. Figures 5 and 6 respectively show networks of term co-occurrence of these two sets of analyses. In both maps, the size of node visualisation associated with each term is proportional to the frequency of occurrence of that term in keyword lists of 'Transportation' articles or in their title and abstracts (depending on which map is considered). Terms that frequently co-occur, i.e. occur jointly in the same article, are generally visualised in closer proximity and form clusters. Both maps are supplemented with hybrid maps, at smaller size representing average year of publication (left) and average number of citations (right) of their articles of origin.

Interestingly, both analyses resulted in only four major clusters of frequently co-occurred terms. This points to the presence of four major divisions of transportation studies. A closer inspection of the content of individual clusters reveals that one cluster is associated with studies of "network analysis and traffic flow" (shown red), one cluster predominantly loaded with terms associated with "economics of transportation and logistics" (shown purple), one predominantly associated with "travel behaviour" studies (shown blue) and one predominantly associated with "road safety" studies (shown orange). Table 1 lists the most frequent terms associated with each cluster, extracted from keywords lists as well as titles and abstract of their underlying articles. These terms provide an abstract characterisation of the nature and diversity of topics of studies within each major division of transportation research.

**Table 1** Most frequent terms associated with each major cluster of 'Transportation' publications.

| ID | Colour | Topic | Top keywords | Further top terms from titles and abstracts |
|---|---|---|---|---|
| #1 | red | **network analysis and traffic flow** | model, optimisation; algorithm; congestion; simulation (model); route choice; networks; network design; (travel) time; information; uncertainty; reliability; constraint; traffic assignment; user equilibrium; route choice; solution; cell transmission; traffic flow | path; congestion; delay; road network; capacity; uncertainty; schedule; traffic condition |
| #2 | purple | **economics of transportation and logistics** | demand; management; competition; impact; cost; price; service (quality); efficiency; productivity; industry; infrastructure; policy; ownership; emission; willingness-to-pay; preferences; market; consumer satisfaction; planning; investment; growth; logistics; supply chain; port; freight transport; airline; shipping; container; carriers; terminal; (transport) policy | company; investment; incentive; electric vehicle; flight; hub; environmental impact; delivery |
| #3 | blue | **travel behaviour** | Travel; travel behaviour; public transport; transit; mobility; accessibility; equity; density; land-use; urban form; city; GIS; residential self-section; mode choice; car ownership; commuting; built environment; environment; perceptions; health; walking; cycling; bicycle; physical activity | Resident; school; neighbourhood; travel pattern; habit; intention; car use |
| #4 | orange | **Road safety** | Behaviour; performance; (road/traffic) safety; risk; (old/young/novice) drivers; attitudes; injury; (injury/crash) severity; statistical analysis; prediction; frequency; heterogeneity; crashes; alcohol; speed; vehicle; pedestrian; children; adults; adolescents; mortality; planned behaviour; fatigue; attention; distraction; (driving) simulator; validation; personality; drinking & sensation seeking | Intervention; questionnaire; score; predictor; crash data; intersection |

Both maps suggest that the cluster of "travel behaviour" is one that represents the youngest articles among the four major clusters. Some of the youngest prominent themes of studies in transportation are represented by the keywords of this cluster, terms such as "built environment" (2016.33)[1], "physical activity" (2016.26), "residential self-selection" (2016.20), "active travel" (2016.88), "active transport" (2016.28) and "walking" (2015.88).

Within the "transport economics and logistics" cluster, the youngest prominent keywords are "electric vehicles" (2016.57), "high-speed rail" (2016.39) "CO2 emissions" (2016.08), "carbon emission" (2015.62) and "infrastructure" (2015.83). In the "traffic assignment and traffic flow" cluster, notable keywords such as "autonomous vehicle" (2018.61), "automated vehicle" (2018.79), "adaptive cruise control" (2017.34), "resilience" (2017.52), and "robustness" (2016.06) are the youngest ones. The road safety cluster is comprised of many old key terms compared to other clusters (terms such as "alcohol" (2010.47) or "drinking" (2010.83)). Among the youngest keywords of the "road safety" cluster, items such as "situation awareness" (2016.26), "statistical analysis" (2016.91), "unobserved heterogeneity" (2016.62), "driver injury severity" (2016.38), "texting" (2016.13), "mobile phone use" (2016.96), "automated driving" (2018.09) and "naturalistic driving study" (2017.48) reflect recurring themes of youngest articles.

The most cited keywords (those whose studies of origin have received the largest counts of citation) within each cluster are "vehicle routing" (48.69)[2], "tabu search" (41.66) and "heuristics" (39.87) (*network and traffic flow* cluster), "mixed logit" (35.31), "supply chain management" (30.43) and "freight transportation" (30.18) (*transport economics and logistics* cluster), "journey" (33.95), "land use" (32.60) and "urban form" (29.36)

[1] Numbers in parentheses present average publication year.
[2] Numbers in parentheses present average number of citations.

(*travel behaviour* cluster), "risky driving" (35.06), "accident risk" (34.08) and "sensation seeking" (30.28) (*road safety* cluster). Generally, most of the keywords whose studies have received the largest counts of citations are concentrated in the *road safety* and *network and traffic flow* clusters compared to the other two clusters (and even more so in the former).

The bibliographic coupling analysis in the previous section identified four distinct clusters of journals with thematically similar articles. One may ask whether any of the four clusters of journals is linked predominantly to any of the four clusters of articles. In order to investigate this question, articles of journals in each cluster were isolated from the rest of 'Transportation' articles and their keywords were analysed separately. Figure 7 shows the outcome in a density view map where hotter colours represent concentration of keywords with higher frequency. Parts (1)-(4) respectively correspond with journal clusters #1 to #4 as identified in Section 4. The analysis does, in fact, indicate a noteworthy association between the clusters of journals and the clusters of article themes as determined by their key terms. Articles of journals that identify with cluster #1 show a high frequency of terms associated with the "travel behaviour" division of articles. Articles of cluster #2 journals present many shared keywords with the cluster representing "economics of transportation and logistics". Cluster #3 journals appear to have published many articles linked to "traffic assignment and traffic flow" division. And cluster #4 journals have predominantly published "road safety" articles. So, while a perfect relation may not exist between clusters of journals and articles (as determined by the co-occurrence of their key terms), there is a close association between the two sets of clusters. This observation further reinforces the idea that the transportation literature, at the most aggregate level, is best characterised by four major divisions.

## 6. Author bibliographic coupling and collaborations in the literature

Authors in the transportation literature were also clustered based on the similarity of article reference lists. Figure 8 presents this bibliographic coupling relationship of authors in a network-view mode where node sizes are proportional to the number of documents from the author as published in 'Transportation' journals. Many authors have published across multiple divisions of the transportation field, so there is no expectation that the clusters match exactly those of the previous four clusters, identified through term co-occurrences, though as we will show, a certain level of association does exist. Here, five major clusters of authors were identified based on the similarity of their referencing.

In order to draw some tangible identity for these clusters on an objective basis, the top ten published authors of each cluster were identified, and their Google Scholar accounts (or if not available, their Scopus author accounts) were examined to record the topics to which they contributed. Based on these self-specified research interests and topics, we were able to determine an overarching label for each cluster of transportation authors. Outcomes have been summarised in Table 2. Although, each author technically identifies with a single cluster when the bibliographic coupling algorithm of VOSviewer is applied, the map shows some authors that are closely associated with more than one cluster. For example, publications of "Wong, S. C." are strongly coupled with publications of authors across both clusters #2 and #4. Similarly, publications of "Bhat, Chandra R." have strong similarities with those of authors across clusters #1 and #4. And a third author whose publications has such feature is "Timmermans, Harry" whose publications are coupled with those of several authors from clusters #1 and #2. The strongest cross-cluster degree of bibliographic coupling in transportation literature is observed between publications of "Meng, Qiang" and "Yang, Hai" followed by publications of "Bhat, Chandra R." and "Eluru, Naveen".

Among authors whose publications are the relatively youngest (i.e. averaging 2017 or later) those with largest numbers of publications within each cluster are "Cats, Oded" (n=51, AY=2017.88); "De Vos, Jonas" (n=31, AY=2017.87) (cluster #1); "Van Arem, Bart" (n=49, AY=2017.18); "Jiang, Rui" (n=30; AY=2017.63) (cluster #2); "Oviedo-Trespalacios, Oscar" (n=25, AY=2018.52); "De Gruyter, Chris" (n=23, AY=2018.00) (cluster #3); "Lee, Jaeyoung" (n=53, AY=2017.71); "Liu, Jun" (n=33, AY=2017.38) (cluster #4); "Wang, Kun" (n=38; AY=2017.61); "Li, Ye" (n=38, AY=2017.33) (cluster #5).

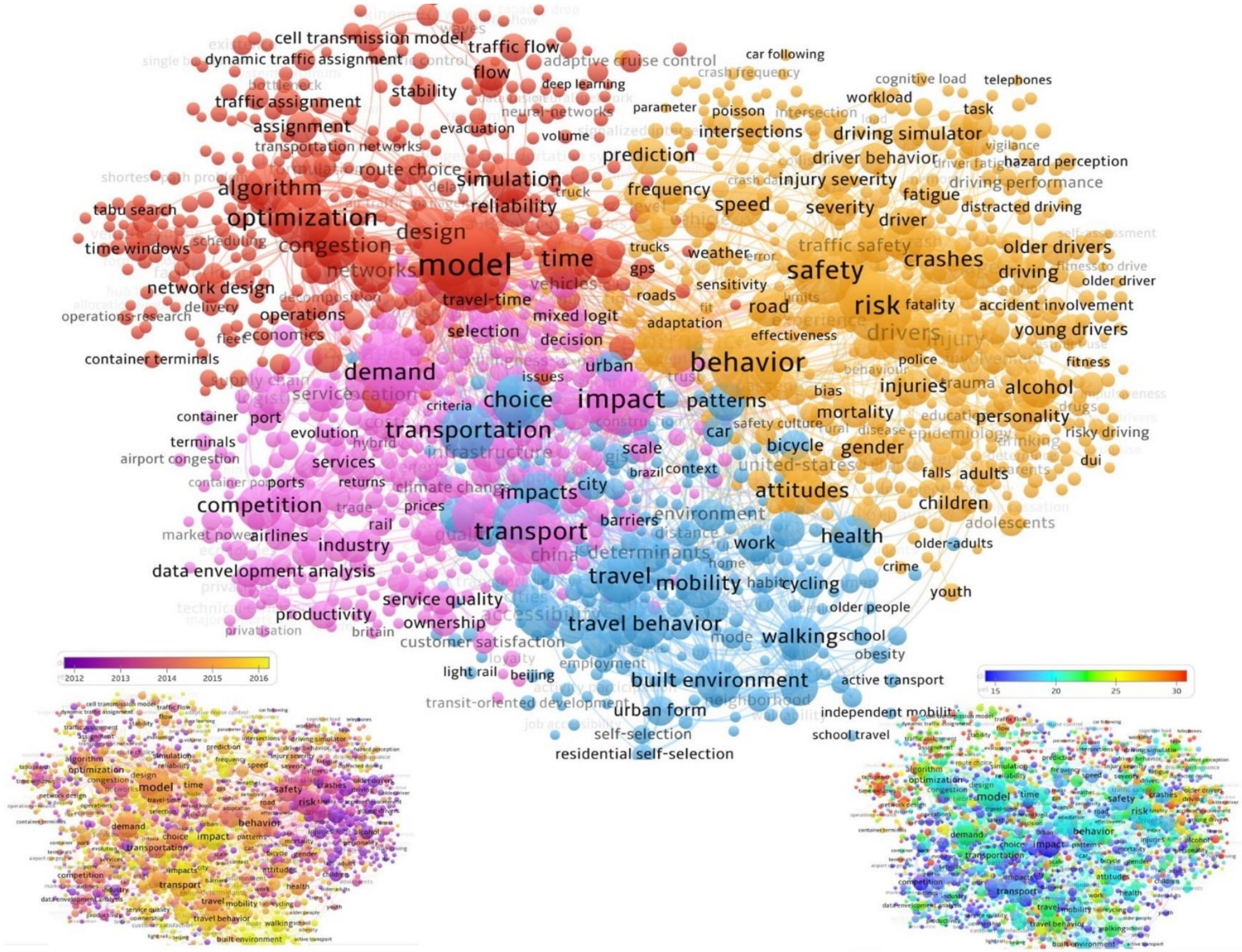

**Figure 5** Keyword co-occurrence in transportation literature. Top left: overlay with average publication year. Top right: overlay with average citation.

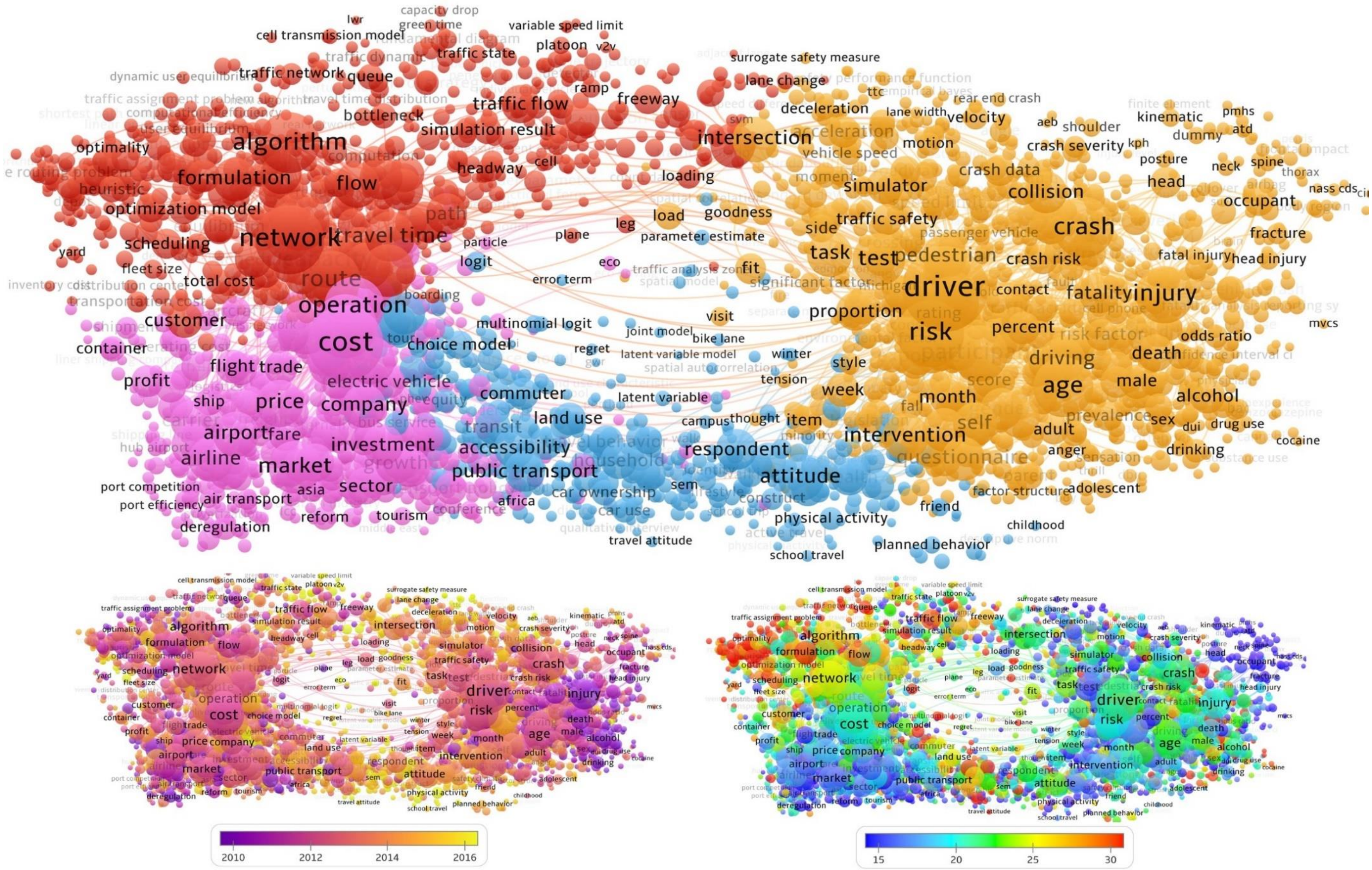

**Figure 6** Term co-occurrence in the title and abstracts of articles in transportation literature. Bottom left: overlay with average publication year. Bottom right: overlay with average citation.

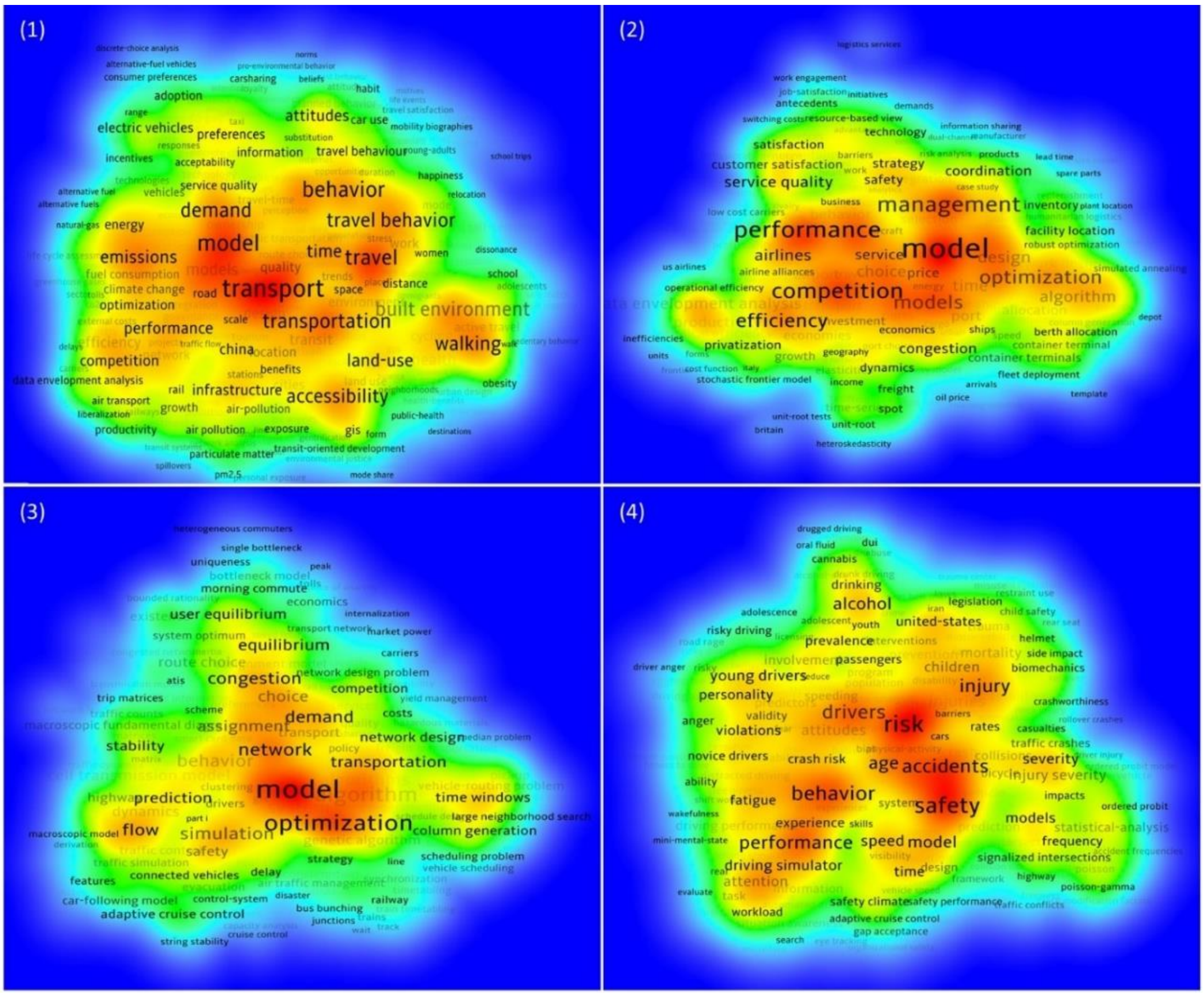

**Figure 7** Maps of keywords of articles of cluster #1, #2, #3 and #4 'Transportation' journals.

**Table 2** Clusters of bibliographically coupled authors in transportation literature.

| Cluster | Label | Ten most published authors (no. of doc.) | Ten most cited authors (no. of citations) |
|---|---|---|---|
| Cluster #1 (grey) n=311 | Travel behaviour, public transport & transport economics | Hensher, David (246); Timmermans, Harry (138); Bhat, Chandra R. (122); Ortuzar, Juan De Dios (89); Van Wee, Bert (89); Axhausen, Kay W. (86); Mokhtarian, Patricia L. (85); Hess, Stephane (82); Currie, Graham (78); Mulley, Corinne (75) | Hensher, David (9757); Bhat, Chandra R. (7198); Mokhtarian, Patricia L. (5921); Axhausen, Kay W. (2881); Kockelman, Kara M (2696); Rose, John M. (2646); Kitamura, R (2439); Golob, TF (2411); Banister, David (2248); Ortuzar, Juan De Dios (2236) |
| Cluster #2 (red) n=237 | Network modelling, optimisation & traffic flow | Yang, Hai (187); Wong, S. C. (165); Lam, William H. K. (131); Daganzo, CF (110); Lo, Hong K. (101); Bell, Michael G. H. (83); Gao, Ziyou (77); Huang, Hai-Jun (76); Szeto, W. Y. (76); Wang, Wei (75) | Daganzo, CF (9621); Yang, Hai (8979); Wong, S. C. (5064); Lam, William H. K. (3762); Lo, Hong K. (3743); Bell, Michael G. H. (3737); Geroliminis, Nikolas (3381); Mahmassani, Hani S. (3067); Huang, Hai-Jun (2738); Yin, Yafeng (2282) |
| Cluster #3 (yellow) n=220 | Traffic psychology & driving behaviour | Williams AF (116); Wets, Geert (72); Prato, Carlo G. (61); Brijs, Tom (59); Lajunen, Timo (57); Watson, Barry (52); Ozkan, Turker (43); Delhomme, Patricia (40); Lenne, Michael G. (38); Brown, Julie (37) | Williams AF (3993); Lajunen, Timo (1961); Watson, Barry (1458); Mccartt, Anne T. (1439); Summala, H (1398); Wets, Geert (1393); Prato, Carlo G. (1084); Lenne, Michael G. (975); Shinar, David (895); Brijs, Tom (887) |
| Cluster #4 (orange) n=161 | Road safety, transportation safety & traffic safety | Abdel-Aty, Mohamed (148); Mannering, Fred (87); Elvik, Rune (76); Eluru, Naveen (73); Sayed, Tarek (70); Huang, Helai (66); Yan, Xuedong (61); Yannis, George (60); Washington, Simon (57); Lee, Jaeyoung (53) | Mannering, Fred (7756); Abdel-Aty, Mohamed (4323); Lord, Domonique (2642); Eluru, Naveen (2332); Washington, Simon (2047); Mohammed A. (1978); Huang, Helai (1715); Elvik, Rune (1403); Wang, Yinhai (1313); Yan, Xuedong (1154) |
| Cluster #5 (green) n=71 | Port and maritime logistics/economics & logistics and supply chain | Meng, Qiang (139); Zhang, Anming (85); Rietveld, Piet (82); Wang, Shuaian (82); Laporte, Gilbert (58); Oum, TH (57); Lam, Jasmine Siu Lee (53); Fu, Xiaowen (50); Sheu, Jiuh-biing (50); Hansen, Mark (49) | Meng, Qiang (4468); Rietveld, Piet (3630); Laporte, Gilbert (3172); Wang, Shuaian (2154); Oum, TH (2062); Zhang, Anming (1582); Sheu, Jiuh-biing (1487); Nikkamp, P (1446); Ropke, Stefan (1227); Lee, Der-horng (1179) |

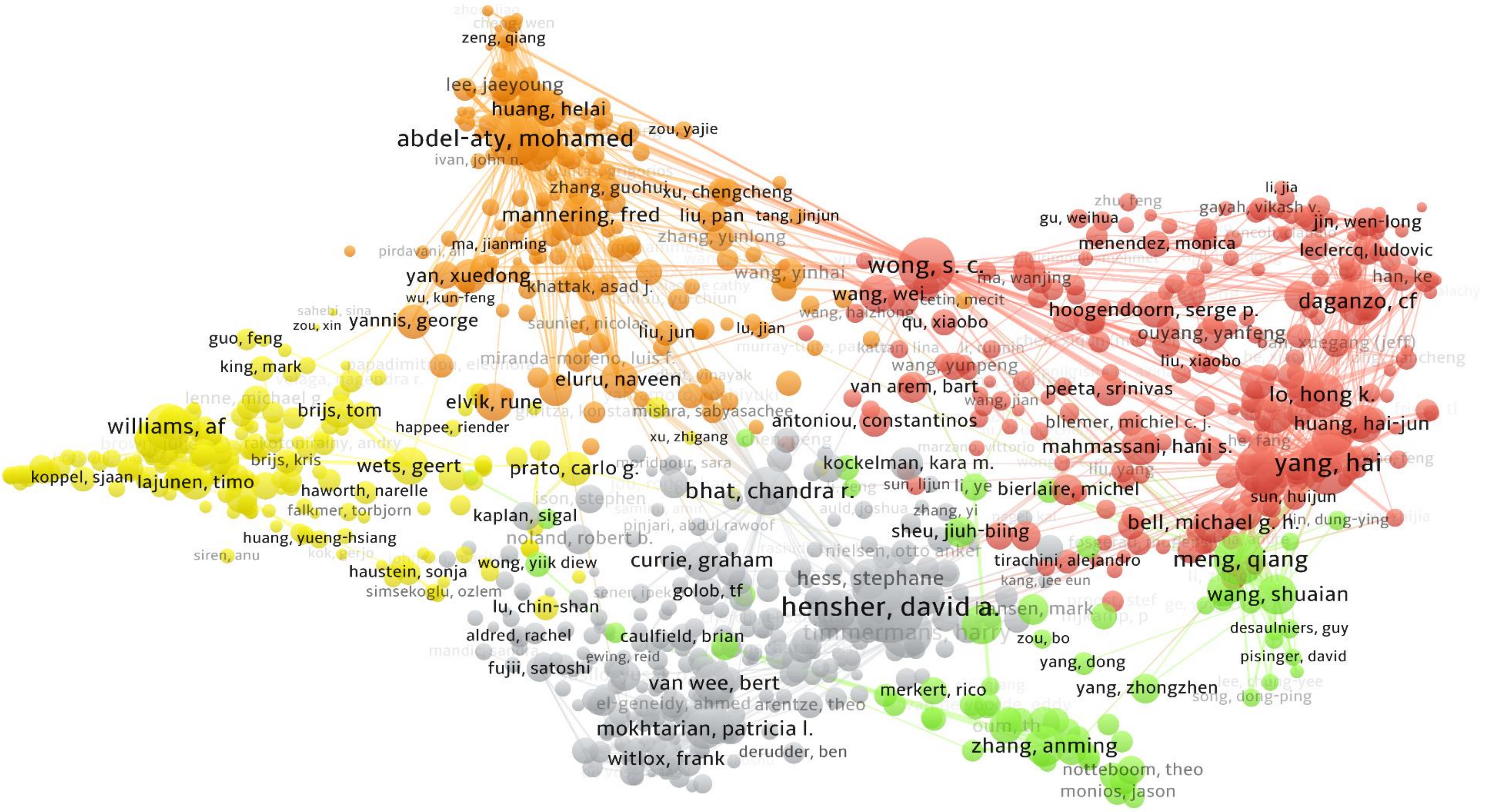

**Figure 8** Bibliographic coupling of authors in transportation literature.

Further analyses were undertaken to identify emerging influential authors of each division. In doing so, authors whose average year of transportation publications were 2017 or later were considered and, within each cluster, authors with largest *normalised citation* count were identified and listed in Appendix C. Some clusters had a higher concentration of emerging authors but in certain clusters, particularly cluster #5 (port and maritime logistics), only a few emerging authors could be distinctly identified.

The network of collaboration between transportation authors is also analysed and is presented in Figure 9 where thickness of the links indicates the frequency of co-authorships between authors. In order to produce a map with sufficient clarity, the analysis was limited to authors with a minimum of 25 transportation documents and minimum of 250 citations to such documents. A total of 21 clusters of co-authorships were identified. The strongest links of co-authorships in transportation literature (i.e., the largest *link strengths* (LS)) is identified between the following pairs of authors: “Brijs, Tom” and “Wets, Geert” (LS=45); “Abdel-Aty, Mohamed” and “Lee, Jaeyoung” (LS=44); “Hensher, David A.” and “Rose, John M.” (LS=37); “Meng, Qiang” and “Wang, Shuaian” (LS=33) and “Timmermans, Harry” and “Rasouli, Soora” (LS=32). Also, authors with strongest degrees of collaborations with other authors within this set of authors are (i.e.; those with largest *total link strengths* (TLS)): “Wong, S. C.” (TLS=138); “Yang, Hai” (TLS=131); “Hensher, David A.” (TLS=119); “Abdel-Aty, Mohamed” (TLS=117); “Meng, Qiang” (TLS=117); and “Timmermans, Harry” (TLS=106). Node sizes of Figure 9 are proportional to the total link strength of each author. Often, authors with larger number of publications have larger total link strengths too. A metric that can represent the diversity of co-authorships of an author is the *number of links* that originate from each node. This represents the number of major authors (authors on the map) with whom each author has collaborated on their transportation articles. Appendix D lists all authors affiliated with each of the 21 clusters of co-authorships along with their number of links and total link strengths. The network of collaboration in transportation literature aggregated to the level of organisations/institutes can be viewed in the Online Supplementary Material.

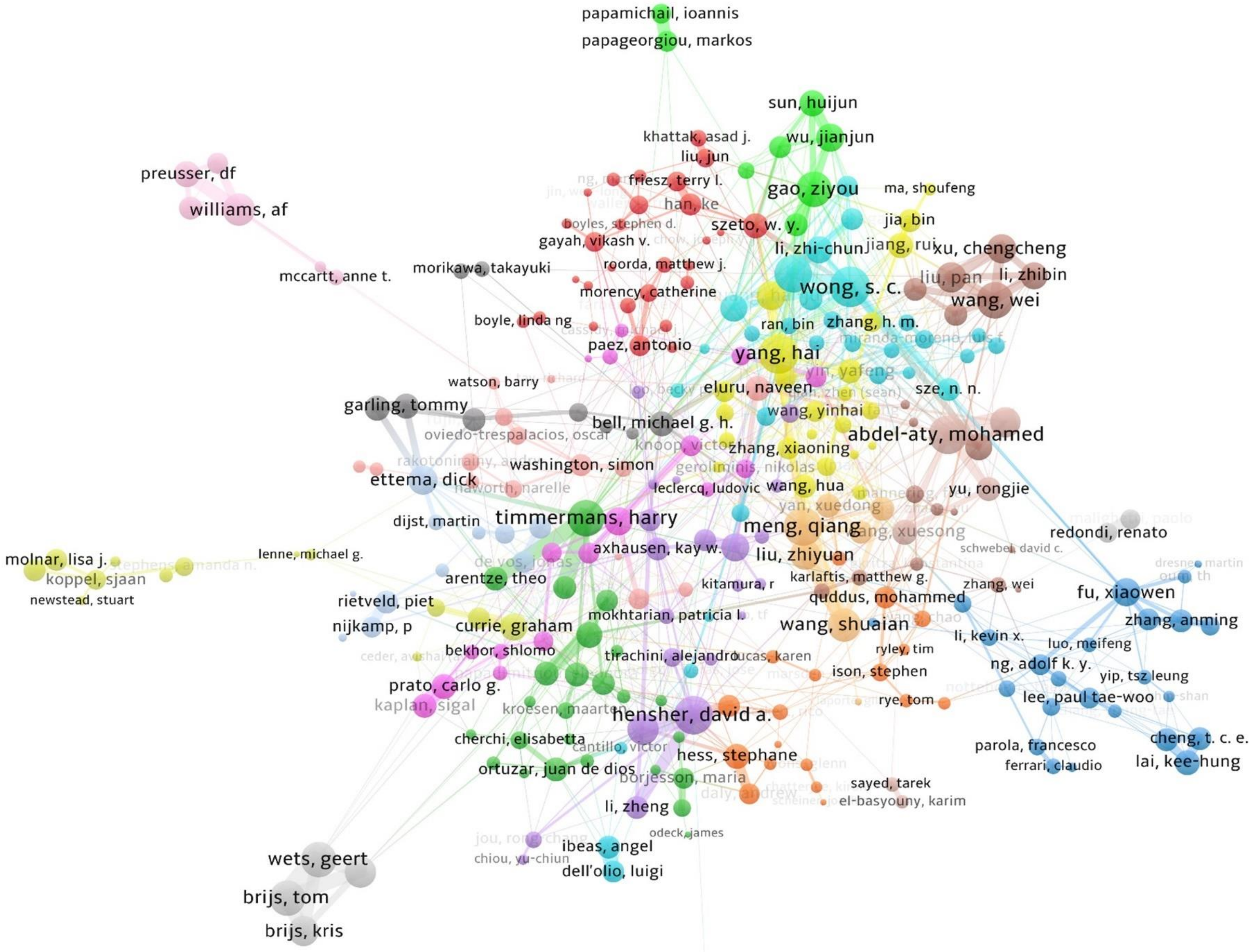

**Figure 9** Co-authorships in transportation literature. Node sizes are based on total link strength.

## 7. Temporal trends, most influential studies and article co-citation patterns in transportation literature

### *7.1. The notions of co-citation, global and local citation, and citation bursts*

In order to provide a finer-grained level of insight into different streams of transportation research (compared to the previous macro-level analyses) and to identify temporal patterns, also in order to identify the most influential studies within each stream of this research, we applied the ground-breaking article co-citation methodology of Chen (2004) for visualising knowledge domains, as implemented in CiteSpace (Chen, 2006). Co-citation is a form of document coupling defined as the frequency with which two documents are cited together by third publications (Boyack and Klavans, 2010; Haghani and Varamini, 2021; Small, 1973).

Consider two documents A and B conceptually illustrated in Figure 10. If article X cites both A and B, this qualifies as an instance of co-citation of documents A and B. The more such citing documents exist (like Y and Z), the stronger the co-citation relationship is between A and B, and the more likely that they are strongly coupled and thematically similar. Articles A and B are called *cited articles* and X, Y and Z are referred to as *citing articles*. By this definition, the publication year of the citing articles has to be equal or greater than the publication year of A and B, whichever is the oldest. Groups of articles that are frequently co-cited can form a cluster, and when the size of such cluster is substantial enough, it could represent a stream of research activity within a field. In the case of our study, each citing article is necessarily a transportation article, i.e. one of the 49,543 articles collectively published by 'Transportation' journals. The cited articles, however, could include publications outside the transportation domain, although they will predominantly reside within transportation. The methodology of Chen (2004) "derives a sequence of co-citation networks from a series of equal-length time intervals". Here, the length of each interval is a year. These networks could subsequently be visualised in a panoramic or time-line view where intellectually significant articles can be visualised based on visually salient features. In reference to Figure 10, if we assume that X, Y and Z are all transportation articles, then each citation from them to A or B is a *local citation* from within the field of transportation, distinct from a general count of citation that A or B might have received globally from any article (i.e., transportation and non-transportation articles combined); i.e. *global citation* count.

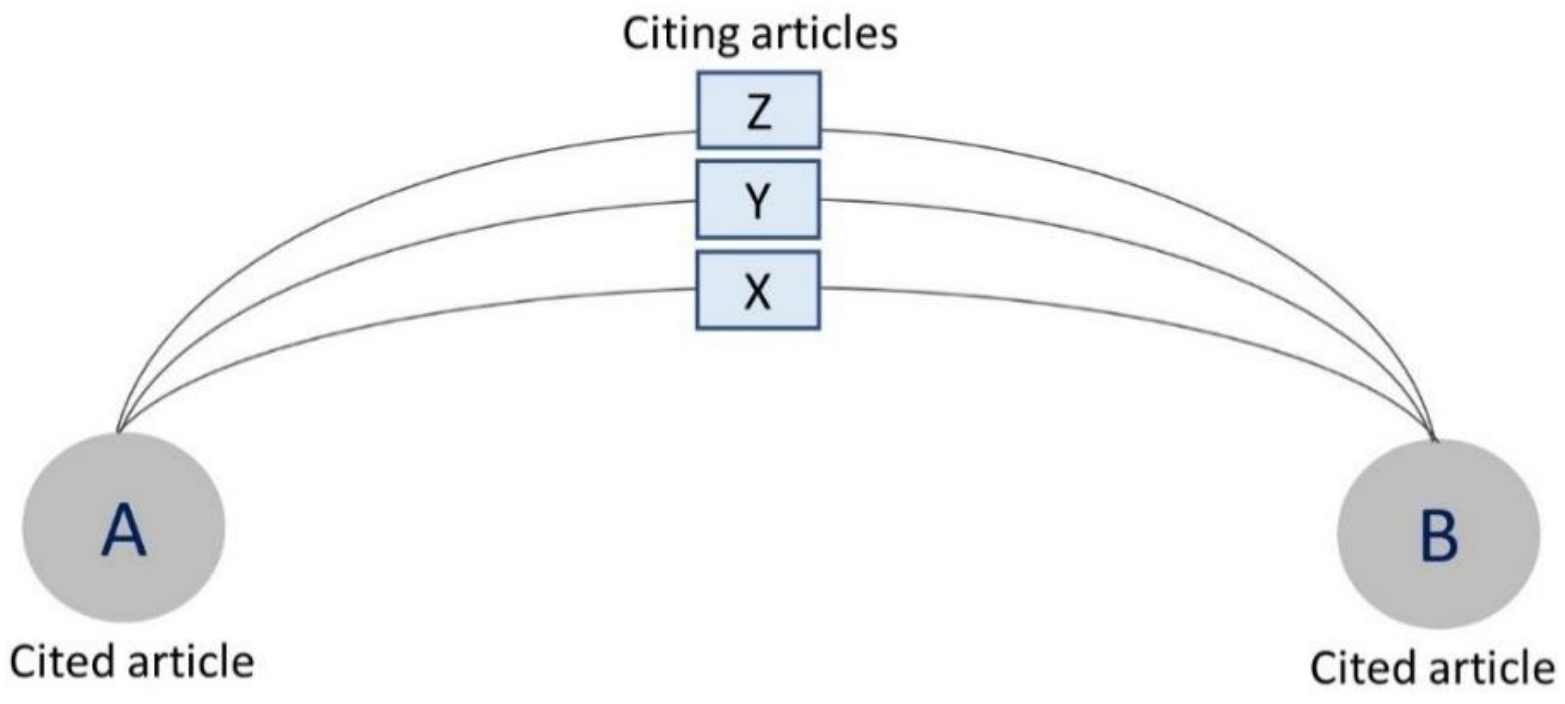

**Figure 10** Illustration of the notion of document co-citation.

The visualisation of articles in a co-citation map is such that the size of the node representing an article is proportional to the number of local citations to that article, as a measure of the impact of that article within the field of interest, here, transportation. Appendix E lists the 30 articles with the largest global citations count, Appendix F lists the top three most cited articles of each 'Transportation' journal based on their global citation count, and Appendix G lists the most cited review articles.

Our analyses, from this point onwards, will exclusively focus on citations from within the field, that is, local citations. Another metric of impact that carries a temporal component is the rate at which (local) citations to individual documents accumulate. Substantial surge in the rate of citation to a document is referred to as *citation burst* and could be an indicator of emerging or hot topics during a certain period of time. Here, the algorithm of Kleinberg (2003) as implemented by CiteSpace is employed for identifying citations bursts.

Articles to which a citation burst has been recorded could also be saliently visualised in a network map of document co-citation. Here (in Figures 11 and 12), red circles represent articles with bursts of citation. A burst of citation is defined by three characteristics: the begin year, the end year (and thereby, a duration), and the strength of burst. Setting a minimum of three years for the duration of bursts, a total of 779 references with citation bursts were identified. Table 3 and Table 4 list the 10 articles in transportation (books excluded) with the strongest and longest citation bursts, respectively. Table 5 also lists the top 20 (books excluded) articles of transportation literature that have received the largest counts of citation from other transportation documents; i.e. most cited within the field. Noting that these cited references could be those published by 'Transportation' journals or otherwise, we use these outcomes to establish *influential outsiders* of transportation literature; i.e., references published by non-transportation journals that have been frequently cited by articles published by 'Transportation' journals. We mark such references with a † sign next to the author's name in tabulations throughout these analyses. Interestingly, most of the highly cited references of transportation literature are, in fact, found to be outsider articles. Similarly, three references with the longest burst of citation in transportation (Peltzman, 1975; Small, 1982; Williams, 1977) are outsider articles. These outsiders will be further analysed in Section 8 after presenting outcomes of clustering for co-cited articles.

**Table 3** References with longest bursts of citation in transportation literature.

| **Author(s) (year)** | **Title** | **Journal** | **Begin** | **End** | **Dur. (yrs)** | **Strength** |
|---|---|---|---|---|---|---|
| Peltzman (1975)† | The effects of automobile safety regulation | Journal of Political Economy | 1977 | 2009 | 33 | 14.7001 |
| Baker et al. (1974)† | The injury severity score: a method for describing patients with multiple injuries and evaluating emergency care | Journal of Trauma and Acute Care Surgery | 1981 | 2011 | 31 | 18.3435 |
| Williams (1977)† | On the formation of travel demand models and economic evaluation measures of user benefit | Environment and Planning A | 1980 | 2008 | 29 | 26.899 |
| Manski (1977)† | The structure of random utility models | Theory and Decision | 1982 | 2009 | 28 | 7.1597 |
| Evans (1976) | Derivation and analysis of some models for combining trip distribution and assignment | Transportation Research | 1980 | 2006 | 27 | 22.4195 |
| Walters (1961)† | The Theory and Measurement of Private and Social Cost of Highway Congestion | Econometrica | 1986 | 2010 | 25 | 17.4472 |
| Fisk (1980) | Some developments in equilibrium traffic assignment | Transportation Research Part B: Methodological | 1982 | 2005 | 24 | 14.4249 |
| Abdulaal and LeBlanc (1979) | Continuous equilibrium network design models | Transportation Research Part B: Methodological | 1983 | 2005 | 23 | 5.8058 |
| Mannering and Winston (1985)† | A dynamic empirical analysis of household vehicle ownership and utilization | The RAND Journal of Economics | 1987 | 2009 | 23 | 13.2412 |
| Fisher (1981)† | The Lagrangian relaxation method for solving integer programming problems | Management Science | 1989 | 2010 | 22 | 6.6206 |

**Table 4** References with strongest bursts of citation in transportation literature

| Author(s) (year) | Title | Journal | Begin | End | Dur. (yrs) | Strength |
|---|---|---|---|---|---|---|
| Mannering et al. (2016) | Unobserved heterogeneity and the statistical analysis of highway accident data | Analytic Methods in Accident Research | 2017 | 2020 | 4 | 49.7725 |
| Smith (1979b) | The existence, uniqueness and stability of traffic equilibria | Transportation Research Part B: Methodological | 1982 | 2001 | 20 | 38.4623 |
| Small (1982)† | The scheduling of consumer activities: work trips | The American Economic Review | 1986 | 2012 | 27 | 36.9685 |
| LeBlanc et al. (1975) | An efficient approach to solving the road network equilibrium traffic assignment problem | Transportation Research | 1979 | 1995 | 17 | 32.3890 |
| Dafermos (1980) | Traffic Equilibrium and Variational Inequalities | Transportation Science | 1982 | 2002 | 21 | 31.5640 |
| Zohar (1980)† | Safety climate in industrial organizations: theoretical and applied implications | Journal of Applied Psychology | 2003 | 2013 | 11 | 31.1641 |
| Chen et al. (2000)† | Carrying Passengers as a Risk Factor for Crashes Fatal to 16- and 17-Year-Old Drivers | JAMA | 2003 | 2011 | 9 | 29.1020 |
| Williams (2003) | Teenage drivers: patterns of risk | Journal of safety research | 2006 | 2014 | 9 | 28.8703 |
| Mannering and Bhat (2014) | Analytic methods in accident research: Methodological frontier and future directions | Analytic Methods in Accident Research | 2016 | 2020 | 5 | 28.5210 |
| Hensher and Greene (2003) | The Mixed Logit model: The state of practice | Transportation | 2005 | 2013 | 7 | 27.9027 |

**Table 5** Top locally cited references in transportation literature

| Author(s) (year) | Title | Journal | LCC |
|---|---|---|---|
| Ajzen (1991)† | The theory of planned behavior | Organizational Behavior and Human Decision Processes | 721 |
| Ewing and Cervero (2010)† | Travel and the built environment: a meta-analysis | Journal of the American Planning Association | 511 |
| Richards (1956)† | Shock Waves on the Highway | Operations Research | 455 |
| Vickrey (1969)† | Congestion theory and transport investment | The American Economic Review | 420 |
| Lord and Mannering (2010) | The statistical analysis of crash-frequency data: A review and assessment of methodological alternatives | Transportation Research Part A: Policy and Practice | 412 |
| Cervero and Kockelman (1997) | Travel demand and the 3Ds: Density, diversity, and design | Transportation Research Part D: Transport and Environment | 400 |
| Daganzo (1994b) | The cell transmission model: A dynamic representation of highway traffic consistent with the hydrodynamic theory | Transportation Research Part B: Methodological | 388 |
| Reason et al. (1990)† | Errors and violations on the roads: a real distinction? | Ergonomics | 345 |
| Lighthill and Whitham (1955)† | On kinematic waves II. A theory of traffic flow on long crowded roads | Proc. of the Royal Society of London. Series A. Mathematical and Physical Sciences | 330 |
| Daganzo (1995b) | The cell transmission model, part II: Network traffic | Transportation Research Part B: Methodological | 308 |
| Small (1982)† | The scheduling of consumer activities: work trips | The American Economic Review | 303 |
| Wardrop (1952)† | Road paper. some theoretical aspects of road traffic research | Proceedings of the institution of civil engineers | 299 |
| Charnes et al. (1978)† | Measuring the efficiency of decision making units | European Journal of Operational Research | 298 |
| Geurs and van Wee (2004) | Accessibility evaluation of land-use and transport strategies: review and research directions | Journal of Transport Geography | 294 |
| Kahneman (1979) | Prospect theory: An analysis of decisions under risk | Econometrica | 291 |
| McFadden and Train (2000)† | Mixed MNL models for discrete response | Journal of Applied Econometrics | 264 |
| Hu and Bentler (1999)† | Cutoff criteria for fit indexes in covariance structure analysis: Conventional criteria versus new alternatives | Structural Equation Modeling: A multidisciplinary journal | 255 |
| Mannering and Bhat (2014) | Analytic methods in accident research: Methodological frontier and future directions | Analytic Methods in Accident Research | 252 |
| Steg (2005) | Car use: lust and must. Instrumental, symbolic and affective motives for car use | Transportation Research Part A: Policy and Practice | 248 |
| Aarts and van Schagen (2006) | Driving speed and the risk of road crashes: A review | Accident Analysis and Prevention | 240 |

### *7.2. The network and visualisation of document co-citation*

Outcomes of the document co-citation analysis, and the respective clusters of co-cited documents, have been visualised in Figure 11, while a corresponding timeline view visualisation is presented in Figure 12. The time span for this analysis has been set to 1970-2020, and the number of look-back years is unlimited, meaning that when analysing each citing article, every single reference cited by each article has been considered regardless of how old the reference has been compared to the publication date of the citing article itself. The criterion for the selection of articles for visualisation is *g-index* (Egghe, 2006) as the standard and default criterion of CiteSpace. The duration of time slices is one year. Each node represents a selected cited article while links indicate the existence of notable co-citation relation between articles. The size of each node is proportional to the number of local citations to the article that it represents. Note that for the purpose of clarity of visualisation, articles labels have not been shown individually (except for the top two most locally cited items of each cluster, in Figure 11). An overlay of a red circle means that the article has also a burst of citation recorded to it. The size of such red circle is proportional to the duration of the burst. The networks presented in Figures 11 and 12 each include 3,392 nodes and 17,324 links. Links are colour-coded in a way that it reflects the average year of the cited references of the cluster that they belong to, with darker colours representing older clusters.

The clustering algorithm of CiteSpace has been applied to identify the clusters of co-cited references. In total, 22 major clusters were identified. Note that the exact count of such clusters does not matter, as there may be several small and insignificant clusters that are outputted too. Each cluster is assigned an identification number, and smaller numbers generally represent more populated clusters. Note that some clusters may be found too small and therefore insignificant and as a result, may not be visualised or receive a label or identification number. Hence, as the identification numbers increase, some discontinuity could be detected in the list of clusters. In order to give certain degrees of insight into the nature/theme of each cluster, Citespace assigns a label by applying a *log-likelihood (LL) ratio* algorithm to noun phrases extracted from the titles of the citing articles of each cluster. If a noun phrase, for example, is repeated in title of several citing articles of a cluster, especially those that have cited many of the references of that cluster (i.e., citing articles with highest *coverage* of the references of the cluster), it will then be likely that the phrase be chosen as the *label* of that cluster. While such labelling is a standard way of assigning an abstract and objectively determined descriptor to clusters, it should be noted that these labels are merely a highly abstract identification and do not specify the exact theme of the entire cluster. Therefore, one should not read too much into the cluster label per se, rather, a series of terms with highest LL scores should be considered in relation with each cluster in order to have a more accurate insight about the nature of the cluster. For example, a label such as *driving anger* (cluster #1) only indicates that the cluster is representative of a road safety-related stream of research, while it may contain a variety of topics within itself, as opposed to being exclusively about driving anger. In this specific case, the noun phrase *driving anger* has been detected in the title of 12 individual citing articles of cluster #1 (articles that also display a relatively high coverage of the cited items of this cluster in their list of references). Therefore, this noun phrase has received the highest LL ratio score among other noun phrases of the citing articles of this cluster and has been chosen as the label/descriptor of the cluster. However, the cluster itself represents a variety of topics such as *young driver* (n=23); *driving behaviour* (n=20); *risky driving* (n=16); *aggressive driving* (n=7); *teenage driver* (n=6); *accident involvement* (n=3). The numbers within parentheses show the number of citing articles of this cluster in which the noun phrase has occurred. Therefore, all one could infer from this set of terms is that this cluster represents sub-topics of driving behaviour within a broader division of road safety studies.

The quality of the clustering in the network is assessed by two common metrics: the *Modularity Q* and the *Silhouette Score* (*SS*). In both cases, values closer to 1 represent a better clustering quality. In the case of the network representing the transportation literature, the Modularity Q is 0.8391 which is a relatively high number. The SS associated with each cluster has been reported in Table 6 along with the *Size* (S) and *Mean Year* (MY) of the cited references of each cluster, top terms and most influential references of each cluster. The list of most influential items includes references with strongest citation bursts as well as those that are highly cited but may not have had a burst of citation. This is to ensure that highly influential articles of each cluster that have acquired their citations steadily (without any burst) are not missed. There is no absolute criterion or cut-off point for identifying such articles, considering that the average citation counts vary significantly across clusters. Some clusters are composed of many highly cited articles whereas the average of citation count for another cluster could be much smaller. Therefore, influential items should be regarded as such within the stream of research that their respective cluster represents. The protocol that was followed includes sorting the cited references based on the strength of the citation bursts, set a cut-off point within the cluster and after that, searching for items with no burst or small burst but very large citation count in the cluster. Most often, items with strong bursts are also highly cited items of the cluster too, but exceptions do exist, and the abovementioned protocol ensures that such influential items are identified. Table 6 also lists citing articles of each cluster that have had the highest amount of coverage of the references of that cluster. These are articles that have contributed the most to the creation of that cluster. These items could be more recently published articles with not many citations recorded to them. The list could be a good indicator of whether the cluster is alive or largely extinct. When top high coverage citing articles of a cluster are all items dating back to 1980s and 1990s (e.g. cluster #3), for example, this is a sign that the cluster is no longer active. On the contrary, a cluster whose high coverage citing articles are all 2019 and 2020 items could represent a likely hot topic.

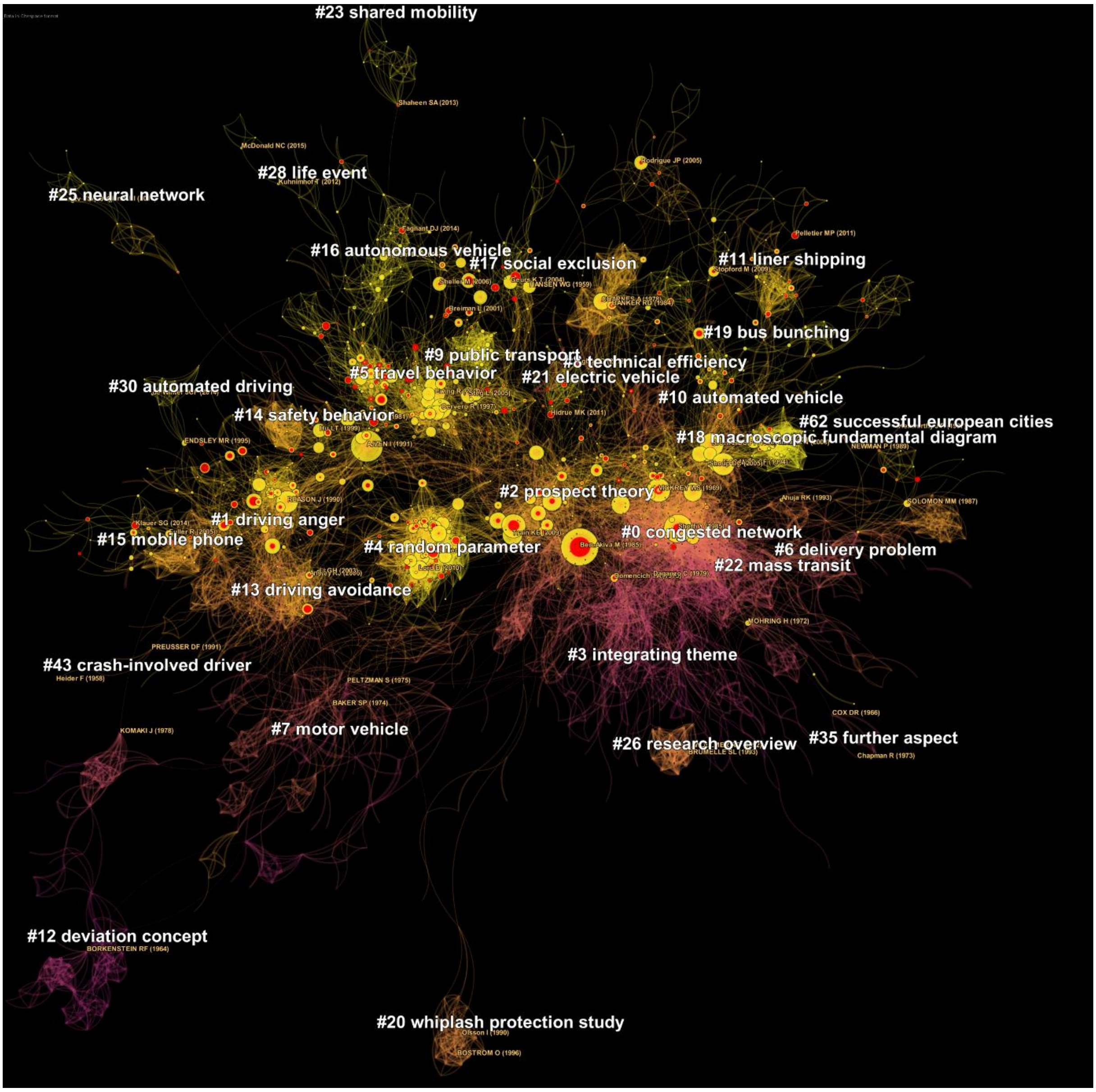

**Figure 11** Network of document co-citation of transportation literature.

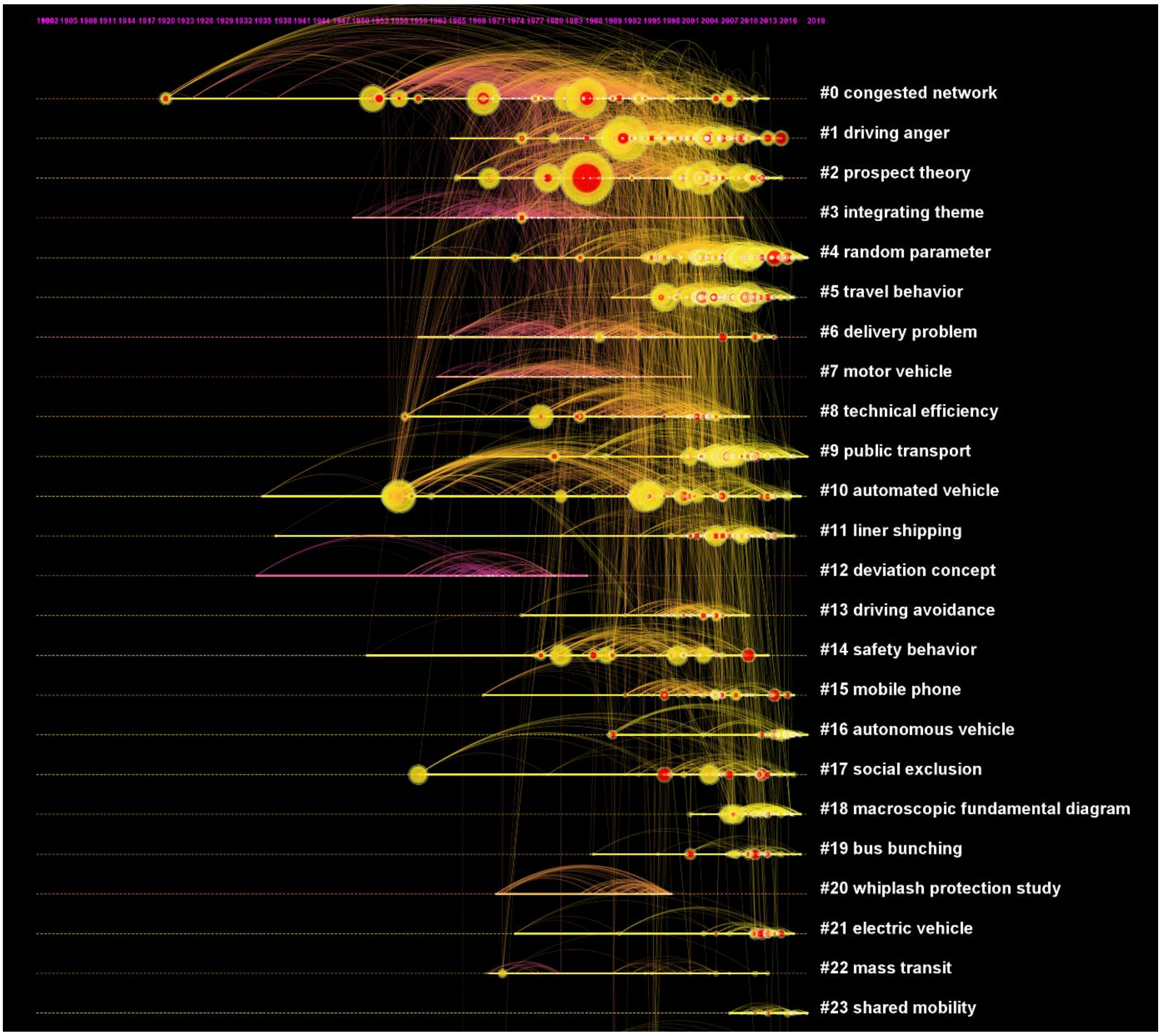

**Figure 12** Timeline view network of document co-citation of transportation literature.

**Table 6** Clusters of co-cited articles in transportation literature. † and * respectively indicate books and outsider articles of transportation literature.

| ID | Size (S), Mean Year (MY), Silhouette Score (SS) | Label and top terms based on LL ratio | Most influential cited references of the cluster | Citing articles with highest coverage of the cluster |
|---|---|---|---|---|
| 0 | S=374<br>MY=1983<br>SS=0.820 | **congested network**; transit network; road pricing; elastic demand; origin-destination matrices; traffic assignment problem; route guidance; bottleneck congestion; modal split; route choice; user equilibrium | Vickrey (1969)†; Beckmann et al. (1955)*; Sheffi (1985)*; Small (1982)†; Wardrop (1952)†; Dial (1971); Daganzo et al. (1977); Pigou (2013)*; Small et al. (2007)*; Arnott et al. (1993)†; Arnott et al. (1990)†; Yang and Bell (1998); Smith (1979a); LeBlanc et al. (1975); Dafermos (1980) | Liu and Szeto (2020) |
| 1 | S=267<br>MY=1997<br>SS=0.878 | **driving anger**; young driver; driving behaviour; risky driving; teenage driver; driving experience; accident involvement; aggressive driving; driver licensing; planned behaviour | Ajzen (1991)†; Evans (1991)*; Williams (2003); Jonah (1997); Deery (1999); Parker et al. (1995)†; Elander et al. (1993)†; Ulleberg and Rundmo (2003)†; Deffenbacher et al. (1994)†; Armitage and Conner (2001)†; Reason et al. (1990)†; Aarts and van Schagen (2006); Cohen (1988)* | Falco et al. (2013); Roidl et al. (2013) |
| 2 | S=224<br>MY=1996<br>SS=0.882 | **prospect theory**; stated preference; stated choice survey; discrete choice model; | Ben-Akiva and Lerman (1985)*; Train (2003)*; Louviere et al. (2000)*; Hensher and Greene (2003); Bierlaire | Guo et al. (2020); van de Kaa (2010a); Van De Kaa (2010b) |
| | | activity-based travel analysis; choice behaviour strategies | (2003); McFadden and Train (2000)†; Bates et al. (2001); Small et al. (2005)†; de Dios Ortúzar and Willumsen (2011)*; Brownstone and Small (2005); Kahneman (1979)†; Hensher et al. (2005)*; Bhat (2001); Greene and Hensher (2003); Lam and Small (2001) | |
| 3 | S=209 MY=1975 SS=0.948 | **integrating theme**; response analysis; multinomial probit model; panel analysis; Dutch mobility panel; car ownership decision | Domencich and McFadden (1975); Daganzo (1979)*; Hausman (1978)†; Mannering and Winston (1985)†; DeSerpa (1971)†; | Williams and de Dios Ortuzar (1982); Van Wissen and Meurs (1989); Kitamura (1990) |
| 4 | S=186 MY=2006 SS=0.971 | **random parameter**; injury severity; unobserved heterogeneity; rear-end crashes; statistical analysis; probit model; mixed logit model; driver-injury severities | Lord and Mannering (2010); Mannering et al. (2016); Washington et al. (2020)*; Mannering and Bhat (2014); Lord et al. (2005); Hauer (1997)*; Savolainen et al. (2011); Milton et al. (2008); Eluru et al. (2008); Abdel-Aty and Radwan (2000); Anastasopoulos and Mannering (2009); Spiegelhalter et al. (2002)†; Bhat (2003) | Hamed and Al-Eideh (2020); Al-Bdairi et al. (2020); Islam and Mannering (2020) |
| 5 | S=176 MY=2006 SS=0.956 | **travel behaviour**; residential self-selection; transit-oriented development; land use; physical activity; travel mode choice; travel attitude; bicycle facilities; car ownership; environment effect | Ewing and Cervero (2010)†; Cervero and Kockelman (1997); Saelens and Handy (2008)†; Woodcock et al. (2009)†; Elvik (2009); Van Acker and Witlox (2010); Saelens et al. (2003)†; Jacobsen (2015); Pucher et al. (2010)†; Kitamura et al. (1997); Heinen et al. (2010); Golob (2003); Pucher and Buehler (2008); Handy et al. (2005); Ewing and Cervero (2001); Cao et al. (2009a); Mokhtarian and Cao (2008) | Oppong-Yeboah and Gim (2020); Wang et al. (2020) |
| 6 | S=175 MY=1986 SS=0.957 | **delivery problem**; routing problem; dynamic pickup; exact algorithm; scheduling problem; city logistics; routing model; tree network; freight transport system | Bodin (1983)†; Bektas and Laporte (2011); Ahuja et al. (1993)*; Ropke and Pisinger (2006); Solomon (1987)† | Daskin (1985); Mitrović-Minić et al. (2004); Harker (1985) |
| 7 | S=159 MY=1982 SS=0.956 | **motor vehicle**; psychosocial factor; safety belt law; pre-school children; Honolulu motor vehicle accident; helmet effectiveness | Baker et al. (1974)†; Evans (1986); Peltzman (1975)† | Viano et al. (1989); Webb et al. (1988) |
| 8 | S=125 MY=1995 SS=0.957 | **technical efficiency**; airline industry; productivity change; technical change; productivity growth; efficiency measurement; aircraft noise | Charnes et al. (1978)†; Banker et al. (1984)†; Oum et al. (1992); Gillen and Lall (1997); Borenstein (1989)†; Farrell (1957)† | Haralambides et al. (2010); Wu and Goh (2010); Simões and Marques (2010) |
| 9 | S=105 MY=2006 SS=0.943 | **public transport**; travel satisfaction; travel experience; customer satisfaction; subjective well-being; commute satisfaction; service quality | Steg (2005); Sheller and Urry (2006)†; Banister (2008); Mokhtarian and Salomon (2001); Lai and Chen (2011); Hannam et al. (2006)†; dell'Olio et al. (2010); Hensher et al. (2003); Beirao and Cabral (2007) | Ingvardson et al. (2020); Lades et al. (2020); Wang et al. (2020); Ye et al. (2020) |
| 10 | S=96 MY=1996 SS=0.942 | **automated vehicle**; mixed traffic flow; cooperative adaptive cruise control; autonomous vehicle; connected automated vehicle; car-following model; human driver behaviour | Richards (1956)†; Daganzo (1994a); Lighthill and Whitham (1955)†; Daganzo (1995a); Shladover et al. (2012); Daganzo and Daganzo (1997)*; Van Arem et al. (2006)†; Treiber et al. (2000)†; Newell (1993); Gipps (1981); Newell (2002) | Rashidi et al. (2020); Delpiano et al. (2020); Anderson and Geroliminis (2020); Barmpounakis and Geroliminis (2020) |
| 11 | S=90 MY=2006 SS=0.989 | **liner shipping**; berth allocation; fleet deployment; container liner shipping | Notteboom and Rodrigue (2005); Stopford (2009); Robinson (2002); Meng et al. (2014); Christiansen et al. (2013)† | Lee et al. (2015); Wang et al. (2015b); Wang et al. (2015a) |
| | | network; container port competition | | |
| 12 | S=71 MY=1970 SS=0.993 | **deviation concept**; occupational accident control; cognitive effect; collision involvement | Borkenstein (1964)*; Komaki et al. (1978)† | Moskowitz (1985); Perrine (1973b) |
| 13 | S=61 MY=2001 SS=0.976 | **driving avoidance**; medical condition; licensing authorities; off-road screening test; driver fitness; driver cohort study; driver licensing | Anstey et al. (2005)†; Li et al. (2003); McGwin Jr and Brown (1999); Ball et al. (1998); Baldock et al. (2006) | Langford (2008) |
| 14 | S=55 MY=1995 SS=0.969 | **safety behaviour**; safety behaviour; safety management; safety performance | Hu and Bentler (1999)†; Fornell and Larcker (1981)†; Podsakoff et al. (2003)†; Zohar (1980)† | Jiang et al. (2010) |
| 15 | S=54 MY=2004 SS=0.991 | **mobile phone**; driving performance; cellular phone use; naturalistic driving study; drivers' decision; cognitive distraction | Caird et al. (2008); Fuller (2005); Redelmeier and Tibshirani (1997)†; Klauer et al. (2014)†; McKnight and McKnight (1993); Dingus et al. (2016)†; Brookhuis et al. (1991) | Oviedo-Trespalacios et al. (2020); McCartt et al. (2010) |
| 16 | S=52 MY=2013 SS=0.981 | **autonomous vehicle**; self-driving vehicle; automated vehicle; autonomous electric vehicle; shared automated vehicle | Fagnant and Kockelman (2015); Fagnant and Kockelman (2014); Krueger et al. (2016); Bansal et al. (2016); Kyriakidis et al. (2015); Agatz et al. (2012)†; Davis (1989)† | Narayanan et al. (2020); Eker et al. (2020); Rashidi et al. (2020) |
| 17 | S=49 MY=2006 SS=0.978 | **social exclusion**; high-speed rail; transport disadvantage; transport equity; accessibility; social inclusion; spatial equity | Geurs and van Wee (2004); Hansen (1959)†; Lucas (2012); Handy and Niemeier (1997)†; Páez et al. (2012); Fu et al. (2012); Foth et al. (2013) | Allen and Farber (2020); Kamruzzaman and Hine (2012); Zhao and Cao (2020) |
| 18 | S=48 MY=2012 SS=0.988 | **macroscopic fundamental diagram**; aggregate traffic representation; dynamic user equilibrium | Geroliminis and Daganzo (2008); Daganzo (2007); Daganzo and Geroliminis (2008); Geroliminis et al. (2012); Aboudolas and Geroliminis (2013) | Haddad and Zheng (2018); Guo and Ban (2020); Mohajerpoor et al. (2020); Tilg et al. (2020); Huang et al. (2020) |
| 19 | S=34 MY=2009 SS=0.995 | **bus bunching**; smart card data; spatial heterogeneity; paired-line hybrid transit design; mobile payment | Pelletier et al. (2011); Koushik et al. (2020); Ma et al. (2013); Farahani et al. (2013)† | Jenelius (2020); Chen et al. (2020); Luo and Nie (2020) |
| 21 | S=31 MY=2009 SS=0.994 | **electric vehicle**; charging station; vehicle body type; electric mobility; consumer preference; slow market diffusion | Rezvani et al. (2015); Egbue and Long (2012)†; Hidrue et al. (2011)†; Schuitema et al. (2010); Pearre et al. (2011); Graham-Rowe et al. (2012); Jensen et al. (2013) | Jin et al. (2020); Kim et al. (2020) |
| 22 | S=31 MY=2009 SS=0.994 | **mass transit**; public ownership; metropolitan area; transit subsidies; parking occupancy level; varied parking tariff | Mohring (1972)† | Pucher et al. (1983) |
| 23 | S=27 MY=2015 SS=0.999 | **shared mobility**; free-floating carsharing; internet-based ride-hailing; mode substitution effect; pick-up location | Shaheen and Cohen (2013); Rayle et al. (2016) | Acheampong et al. (2020); Nair et al. (2020); Tirachini et al. (2020); Shen et al. (2020) |

### *7.3. Clusters of co-cited documents in transportation literature and their temporal evolution*

There are readily identifiable associations between aggregate divisions of transportation studies as determined in the previous section through journal and key term analysis and the clusters of activities determined by the document co-citation patterns. Of the various streams of activities detected through the document co-citation clustering, seven clusters identify clearly with the road safety division of transportation research (clusters #1, 4, 7, 12, 13, 14, 15), six clusters are associable with the network modelling and traffic flow division (clusters #0, 6, 10, 18, 19, 23), five clusters are linked predominantly to the transport economics and logistics division (clusters #2, 8, 11, 21, 22) and five clusters are predominantly related to the travel behaviour division (clusters #3, 5, 9, 16, 17). See Appendix H for more details.

Note that each of these individual clusters display different patterns of temporal development, as some are older (i.e., consist of older references with no recent (co)-citation occurrence) and some are younger. Also, activities of some have largely terminated or slowed down while some others are currently considered to be hot research streams in transportation studies. These temporal patterns are best identifiable through a timeline view of the network in Figure 12 along with the dynamic year-by-year development view of the network as presented in an Online Supplementary Video of this article. This video essentially visualises the temporal construction of the document co-citation network from 1970-2020 at time increments of one year. At each time slice (i.e., year), parts of the network (i.e., instances of frequent co-citation) that have occurred during that year are visualised in a visually salient way, allowing an observer to identify the streams that have been most active during each year since 1970. Appendix H also summarises the content of this visualisation in a more abstract way. It displays the periods of time during which each stream has had moderate/intense citation activities. Streams whose references are old (Figure 12) and also have not shown citation activities for several years (Appendix H) can be considered extinct. On the other hand, streams that have young references (Figure 12) as well as several young citing articles (as reflected in the last column of Table 6, or alternatively, Appendix I) constitute active or hot areas.

Cluster #0 *congested network*, for instance, representing a classic stream of transportation research linked predominately to network modelling, seems to be one of the oldest and one of the pioneering streams of activities in this field with references that go as far back as 1950s (Beckmann et al., 1955; Wardrop, 1952). According to the dynamic visualisation, activities of this cluster (i.e., instances of frequent co-citation to the references of this cluster) seem to have started to notably begin since late 1979 and early 1980s. The cluster represents a stream of research that sustained throughout 1980s and constituted the primary stream of transportation research during this period. The seminal book of Sheffi (1985) entitled "*Urban Transportation Networks*" is the single most influential reference among the cited articles of this cluster (n=398). Following the peak of this cluster between 1980 and 1985, clusters related to transport economics, travel behaviour or network modelling, such as cluster #3 *integrating theme*, cluster #8 *technical efficiency*, or cluster #6 *delivery problem* on one hand, and certain road safety clusters such as cluster #7 *motor vehicle* or cluster #13 *driving avoidance*, on the other hand, began to emerge in parallel and started to show notable activities. During early years of the 21st century, two particular clusters seem to show notable and sustained patterns of activities (as reflected in co-citation occurrences) which could be regarded as hot topic at the time. This includes cluster #2 *prospect theory* whose major cited references are predominantly linked to discrete choice modelling studies. Among those references, the seminal book of Ben-Akiva and Lerman (1985) entitled "*Discrete Choice Analysis: Theory and Application to travel demand*", as the single most locally cited reference of transportation research (n=741) has played a fundamental role. A plausible explanation for the sudden invigoration of this cluster since 2000 could be the Nobel Prize of Economics awarded to Daniel McFadden in 2000 (as a pioneering figure of this subdomain) that could have been a catalyst to this. The second hot topic during early years of the 21st century is reflected in activities a road safety cluster, #1 *driving anger*. While activities of this cluster predate 2000 and are detectable since mid-1990s, especially following the seminal publication of Deffenbacher et al. (1994) entitled "Development of a driving anger scale" in *Psychological Reports* (a non-transportation journal), those activities began to intensify during early years of the current century.

Skipping forward to the last five years of the development of the transportation field, we seek to identify the current hot topics of research. A year by year examination of the network discovers a number of hot areas:

(a) One distinct cluster of road safety research, cluster #4 *random parameter*, predominantly focused on statistical analysis of road accident data;
(b) Three clusters of travel behaviour research including cluster #5 *travel behaviour*, predominantly focused on topics of active transport, land use and residential self-selection, cluster #9 *public transport* predominantly related to traveller experience/satisfaction and cluster #17 *social exclusion*, focused predominantly on transport/spatial equity;

(c) One cluster in the network and traffic flow domain, cluster #18 *macroscopic fundamental diagram*.

These areas show patterns associated with hot topics, in that, they have been consistently active almost every year since 2015. On the other side, the co-citation network shows less distinct but younger areas that have not yet proven a sustained level of heightened activity but have shown signs of becoming the emerging hot topics of the field. This is reflected in the activities of clusters #10 & #16 *automated/autonomous vehicle*, and cluster #21 *electric vehicle*. Cluster #23 *shared mobility* also displays patterns of becoming the newest major emerging topic.

## 8. Influential outsiders of transportation literature

Through the document co-citation analyses that were presented previously, we identified a cohort of articles not published by 'Transportation' journals per se, but frequently been cited by articles of 'Transportation' journals. We labelled those as *influential outsiders* of the transportation literature. Appendix I lists these outsider articles along with the division of transportation research that they most closely identify with. This determination has been made based on the content of the article itself, as well as the nature of the citing and cited articles of the cluster with which these outsiders were associated. Note that we have made a distinction here between "economics of transport" and "logistics" and treated them as two different divisions, considering that outsiders of these two sub-divisions were of markedly different content. Also, note that the determination of influential outsiders is exclusively limited to articles and does not include books.

The table in Appendix I shows that the road safety division has the largest number of influential outsiders among all other divisions (n=25). This is followed by the travel behaviour division (n=15). The knowledge foundation of the network and traffic flow studies as well as studies of the logistics division, on the other hand, seem to be relatively more endogenous than other divisions considering that they have much fewer number of influential outsiders. The top cited outsiders of transportation literature are Ajzen (1991) (n=721) (entitled "The theory of planned behavior" in *Organizational Behavior and Human Decision Processes*), Ewing and Cervero (2010) (n=511) (entitled "Travel and the built environment: a meta-analysis" in *Journal of the American Planning Association*) and Richards (1956) (n=455) (entitled "Shock Waves on the Highway" in *Operations Research*). Among non-transportation journals whose items have been found instrumental by transportation studies, *Journal of Applied Psychology* (n=4), *The American Economic Review* (n=4) and *Operations Research* (n=4) have contributed the largest number of items.

## 9. Concluding remarks and directions for future research

### *9.1. Summary and conclusions*

The aim of this study was to visualise and analyse the scientific literature in the transportation field in its approximate entirety based on various scientometric indicators such as the co-occurrence of keywords, similarity of references and frequency of co-citation instances. The focus was on a determining structural composition of the field of transportation at various levels of aggregations and to document its temporal development. The size of this literature is estimated to have grown to nearly 50,000-65,000 articles (depending on whether one considers articles of *Transportation Research Record*), while an excess of 4,000 articles being added to this literature every year (according to the current rate). Since 2007, the rate of publication in the transportation field has exceeded 1,000 items per year and has substantially grown since. A portion of nearly 1.7% of these articles are estimated to be of review/synthesis nature. The record also shows that transportation researchers have shown a heightened level of attention to meta-analytic and scientometric review studies within the last five years. This may be because the rate of publications is so large that reviews are becoming more useful to get an overview of the literature. While meta-analytic studies in earlier years used to be predominantly emerging from the road safety division of this field, the record shows that meta-analyses are being increasingly and very noticeably adopted by authors of other divisions in the recent years (Aston et al., 2020; Bastiaanssen et al., 2020; Govindan et al., 2020; Raza et al., 2020; Semenescu et al., 2020; Singh et al., 2020; Tao and Zhu, 2020).

Analysis of the co-occurrence of terms (extracted from the keywords lists, titles and abstracts of transportation articles) suggested that the literature can be objectively segmented into four major divisions. These divisions are linked to studies of (i) network and traffic flow, (ii) transportation economics and logistics, (iii) travel behaviour and (iv) road safety. This could be regarded as an objective segmentation of this field at the highest degree of aggregation. The analyses on the similarity of referencing of articles at the level of journals largely support this segmentation, although when the analysis is conducted at the level of authors, these divisions slightly change. For example, the road safety cluster splits into two distinct clusters at the level of authors, while transport economic authors merge with those identifying with travel behaviour topics and form their own cluster. The document co-citation analysis pointed clearly to a number of distinct streams of research in this field that have shown consistent, sustained and intense degrees of activities within the last five years and hence, could be regarded as hot topics of the field. This collectively included trending topics such as "land-use", "active transportation", "residential self-selection", "traveller experience/satisfaction", "social exclusion" "transport/spatial equity", "statistical modelling of road accidents" and "macroscopic fundamental diagram". Also, topics of "automated transport", "alternative-fuel vehicles" and "shared mobility" were determined to be younger and emerging trendy topics of the field.

### *9.2. Directions for future research*

When interpreting the findings of this work, particularly when it comes to the streams of transportation research detected by the document co-citation analysis, readers may take the (i) scale and (ii) timing of these analyses into account. Give that analyses are conducted on a full scope of the transportation literature (issue (i)), it is expected that only major streams of this research (as reflected in clusters of co-cited articles) turn up in the outputs. Clearly, some divisions of transportation research are smaller than others (as reflected in the number of authors active in different divisions). For example, the travel behaviour division is much more populated than the division of maritime logistics, and as a result, more research is produced in that area. Therefore, when the literature is analysed in its full scale, streams of maritime logistics research may not necessarily be outputted as major streams of transportation research. This is also the case for topics such as air transportation. While traces of activities of the air transportation branch were detectable in cluster #8 of the document co-citation analyses, this topic does not rise to the scale of a standalone topic, when considered within the full scope of the field. This also has an impact on the citation count of the articles and authors in those smaller divisions and results in giving more visibility to more populated divisions. While this does not detract from the validity of the outcomes (given the intended scale of the analyses), it begs the question of how one can further dissect various divisions of the transportation research in more details and with a higher resolution into their content and activities. The answer is, following the identification of the major divisions of transportation research by this work, it would be warranted that further analyses be dedicated to specific divisions. Research in various sub-divisions of transportation research are also themselves extensive enough to warrant large-scale synthesis of research. This, however, relies on whether the reference dataset of specific sub-divisions can be created and be dissociated from the rest of the body of the literature with acceptable accuracy. For some distinct sub-divisions, such as road safety, the data may be easier to obtain, while for some other sub-divisions, such as travel behaviour whose literature is entangled with the rest of the body of the field, data acquisition may require more creativity and effort in designing a search strategy. The outcomes of the current study could serve as a preliminary guide for authors who may wish to design such search strategies that can isolate various sub-divisions of transportation from the rest of the body of the literature.

It should also be noted that there is often a time lag between the emergence of topics in a field and their rising to a level that can be detected by a large-scale analysis like that of this work (issue (ii), as pointed out above). For example, we detected the topic of shared mobility as one that distinctly shows characteristics of an emerging topic. While activities of this cluster (as reflected in the instances of co-citation) was detectable since 2017, this does not indicate that the topic was only introduced to the literature since 2017. It rather means that the topic began to become a major cluster since that time. It usually takes some time for articles of a certain topic to accumulate and then be co-cited by subsequent papers. This may answer the question of why, for example, some contemporary topics of transportation research such as *mobility as a service* have not left much trace in our outputs. The answer is simply because the topic is very young. According to the WoS, there are 130 articles (at the time of this publication) with the term combination "mobility as a service" in their title. Of these, more than 66% have emerged since 2019 and onwards. Therefore, there has not been enough time for the majority of body of this stream to be (co)-cited by newer articles and to form a cluster detectable at the scale at which we undertook our analyses. An update of these analyse in few years' time may, for example, be able to detect these very young emerging activities as major clusters. Similarly, there are also other important topics of transportation research such as the area known as *pedestrian, crowd and evacuation dynamics* (Haghani, 2020a, b; Haghani and Sarvi, 2018) which did not have a representation in these outcomes. The reason, in this case, is that this is among topics that are not exclusive to the field of transportation and is one whose borders cross several disciplines, including transportation. Such areas would require their own dedicated analysis, as the current study would only detect those that are major mainstream transportation research clusters.

Another dimension to be considered with respect to the identification of influential papers and hot topics is that our study was generally descriptive in those areas. Given the scale of this work, it was not possible to seek causality of why individual papers have become influential or why individual streams have become hot topics. Analyses at smaller scale however (e.g., focused on specific sub-divisions) may also enable us to look at the top influential papers of their respective divisions and ask why those studies rose to such prominence. Same question can be asked in relation to the hot topics. A topic can become trendy in a field of research for a variety of reasons and this reason is not necessarily that there is a societal or scholarly urgency for such research. A Topic may become trendy due to various reasons or their combination such as, (a) publication of a paper on a topic in top journals such as *Nature* or *Science* (an example is the paper of Helbing et al. (2000) in the field of pedestrian/crowd dynamics), (b) the launch of a new specialty journal that can invigorate certain areas of a field (e.g., *Analytic Methods in Accident Research*), (c) major methodological/computational advancements, (d) major technological advancements (e.g., automation in mobility), or (e) intense activities of one (or a few) specific research teams on a certain topic during a certain period of time, that may result in heavy number of publication from those groups on that topic. Research on the causality of hot topics of transportation could be another interesting dimension for a follow-up study.

It is hoped that these findings have provided a new level of insight into the field of transportation research, its journals and its various streams of activities as well as its temporal trends. Documentation of the most influential articles and authors is also expected to assist new researchers of transportation, particularly in undertaking conventional review studies.

## Conflict of interest statement

On behalf of all authors, the corresponding author states that there is no conflict of interest.

## Authors' contributions

**Milad Haghani**: Conceptualization; Data curation; Formal analysis; Investigation; Methodology; Software; Visualization; Writing - original draft
**Michiel C. J. Bliemer**: Conceptualization; Funding acquisition; Investigation; Methodology; Resources; Writing - review & editing

## Acknowledmengts

This research was funded by Australian Research Council grants DP180103718, DP150103299 and DE210101175.

**Appendix A – The search query for acquisition of the reference dataset of transportation literature**

SO=("analytic methods in accident research" OR "transport reviews” OR “transportation research part b-methodological” OR “transportation research part e-logistics and transportation review” OR “transportation research part d-transport and environment” OR “transportation” OR “transportation research part c-emerging technologies” OR “transportation research part a-policy and practice” OR “journal of transport geography” OR “accident analysis and prevention” OR “transportation science” OR “transport policy” OR “travel behaviour and society” OR “journal of intelligent transportation systems” OR “maritime policy & management” OR “journal of safety research” OR “journal of air transport management” OR “international journal of sustainable transportation” OR “transportation research part f-traffic psychology and behaviour” OR “transportmetrica a-transport science” OR “journal of transport & health” OR “european transport research review” OR “transportmetrica b-transport dynamics” OR “research in transportation business and management” OR “research in transportation economics” OR “transportation letters-the international journal of transportation research” OR “maritime economics & logistics” OR “economics of transportation” OR “journal of transportation safety & security” OR “journal of advanced transportation” OR “mobilities” OR “journal of transport and land use” OR “traffic injury prevention” OR “journal of public transportation” OR “transportation journal” OR “european journal of transport and infrastructure research” OR “journal of transport economics and policy” OR “international journal of shipping and transport logistics” OR “international journal of transport economics”)

**Appendix B – The search query for acquisition of the review articles of transportation literature**

TI=(“literature review” OR “literature survey” OR “literature mapping” OR “A review” OR “scoping review” OR “systematic review” OR “systematised review” OR “systematic survey of the literature” OR “comprehensive review” OR “critical review” OR “mapping review” OR “mixed methods review” OR “bibliometric review” OR “scientometric review” OR “evidence synthesis” OR “synthesis of evidence” OR “rapid review” OR “state-of-the-art review” OR “review of the state of the art” OR “qualitative review” OR “quantitative review” OR “umbrella review” OR “meta analysis” OR “meta-analysis” OR “meta-analytic” OR “meta synthesis” OR “meta-synthesis” OR “meta evidence”) OR AK=(“literature review” OR “literature survey” OR “systematic review” OR “scoping review” OR “critical review” OR “literature analysis” OR “literature mapping” OR “bibliometric” OR “scientometric” OR “survey of literature” OR “meta-analysis” OR “meta research”))

**Appendix C –** Clusters of emerging influential authors based on similarity of their references in their transportation publications 2017 or later, and based on normalised citation (min=40.00)

| **Author cluster** | **Dominant topics within each cluster** | **Author name (average publication year, normalised citation)** |
|---|---|---|
| Cluster #1 (grey) | Travel behaviour, public transport & transport economics | De Vos, Jonas (2017,87, 93,45); Wang, Shanyong (2017.75, 83,08); Cats, Oded (2017.88, 90.88); Wang, Jing (2017.36, 74.03); Chen, Peng (2017.32, 63.88); Zhao, Jinhua (2017.41, 52.15); Ding, Chuan (2017.64, 43,46); Wang, Kun (2017.64, 43.46); Rasouli, Soora (2016.91, 40.46); |
| Cluster #2 (red) | Network modelling, optimisation & traffic flow | Wang, Yunpeng (2017.10, 112.02); Menendez, Monica (2017.11, 86.27); Wang, Meng (2017.67, 59.80); Ramezani, Mohsen (2017.60, 56.59); Ma, Jiaqi (2017.86, 54.60); Chen, Xiqun (Michael) (2017.67, 53.18); Wu, Jianjun (2017.50, 52.92); Liu, Yang (2017.12, 50.50); He, Fang (2016.88, 71.60); Sun, Huijun (2017.21, 44.45); He, Zhengbing (2017.47, 45.26); Jiang, Rui (2017.63, 41.16); Saberi, Meead (2017.58, 40.20), Haghani, Milad (2018.15, 40.01); Jia, Bin (2017.01, 41.16) |
| Cluster #3 (yellow) | Traffic psychology & driving behaviour | Oviedo-Trespalacios, Oscar (2018.52, 62.52); Wang, Jing (2017.36, 74.03); Happee, Riender (2017.88, 47.15); Qu, Xingda (2018.78, 42,59); Zhang, Tingru (2017.42, 43.57); |
| Cluster #4 (orange) | Road safety, transportation safety & traffic safety | Lee, Jaeyoung (2017.71, 114.24); Guo, Yanyong (2018.40, 75.71); Wen, Huiying (2018.38, 58.26); Fountas, Grigorios (2018.27, 51.53); Hao, Wei (2018.05, 48.40); Cai, Qing (2018.35, 48.28); Wang, Ling (2017.79, 41.59) |
| Cluster #5 (green) | Port and maritime logistics/economics & logistics and supply chain | Van Woensel, Tom (2017.17, 52.05); Yang, Zaili (2017.19, 40.53); |

**Appendix D –** Clusters of co-authorships in transportation literature for major authors of the field. Number of links represent the number of co-authors on the network while link strengths represent the number of co-authored publications with other authors of the network. Authors of each cluster are listed alphabetically.

| Author cluster | Author name (number of links, total link strength) |
|---|---|
| #1 (n=28) | Boyle, Linda NG (2, 8); Boyle, Linda NG (3, 6); Cherry, Christopher R. (3, 4); Chow, Joseph Y. J. (3, 3); Donnell, Eric T. (1, 5); Dozza, Marco (4, 4); Farber, Steven (4, 14); Friesz, Terry L. (4, 23); Gayah, Vikash V. (9, 22); Guo, Feng (5, 5); Habib, Khandker Nurul (2, 5); Han, Ke (5, 28); Jin, Wen-Long (2, 3); Khattak, Asad J. (3, 16); Kockelman, Kara M. (4, 7); Lee, John D. (2, 8); Liu, Jun (6, 21); Miller, Eric J. (4, 9); Mishra, Sabyasachee (3, 3); Morency, Catherine (4, 18); NG, Manwo (5, 10); Nordfjaern, Trond (1, 2); Paez, Antonio (8, 26); Roorda, Matthew J. (6, 13); Schonfeld, Paul (2, 2); Scott, Darren M. (5, 7); Szeto, W. Y. (16, 41); Waller, S. Travis (9, 21) |
| #2 (n=25) | Antoniou, Constantinos (12, 31); Arentze, Theo (6, 36); Ben-Akiva, Moshe (6, 9); Bierlaire, Michel (6, 6); Borjesson, Maria (7, 26); Carlos Martin, Juan (3, 5); Carlos Munoz, Juan (5, 9); Cherchi, Elisabetta (6, 18); Chorus, Caspar (14, 35); Cirillo, Cinzia (5, 6); Eliasson, Jonas (5, 17); Elvik, Rune (5, 9); Fosgerau, Mogens (4, 6); Handy, Susan (5, 6); Haustein, Sonja (4, 5); Kroesen, Maarten (5, 16); Odeck, James (1, 1); Ortuzar, Juan De Dios (12, 36); Papadimitriou, Eleonora (4, 37); Proost, Stef (3, 5); Rasouli, Soora (4, 35); Timmermans, Harry (21, 106); Van Wee, Bert (17, 36); Walker, Joan L. (5, 7); Yannis, George (7, 46) |
| #3 (n=24) | Chang, Young-Tae (5, 10); Cheng, T. C. E. (5, 35); Dresner, Martin (2, 2); Ferrari, Claudio (3, 13); Fu, Xiaowen (14, 55); Lai, Kee-Hung (5, 39); Lam, Jasmine Siu Lee (14, 18); Lee, Paul Tae-Woo (11, 26); Li, Kevin X. (10, 16); Lu, Chin-Shan (4, 6); Lun, Y. H. Venus (7, 35); Luo, Meifeng (8, 12); NG, Adolf K. Y. (10, 27); Notteboom, Theo (7, 16); Oum, TH (3, 10); Parola, Francesco (3, 15); Vanelslander, Thierry (2, 4); Wang, Kun (8, 30); Wong, Yiik Diew (3, 4); Yang, Zaili (5, 14); Yang, Zhongzhen (7, 17); Yip, Tsz Leung (7, 11); Zhang, Anming (7, 34); Zhang, Yahua (3, 19) |
| #4 (n=23) | Chen, Feng (5, 5); Chen, Peng (4, 17); Gao, H. Oliver (3, 3); He, Fang (6, 18); Huang, Hai-Jun (14, 59); Jia, Bin (12, 33); Jiang, Rui (13, 40); Liu, Wei (6, 31); Ma, Shoufeng (3, 8); Nie, Yu (Marco) (11, 19); Qian, Zhen (Sean) (5, 11); Sheu, Jiuh-Biing (4, 4); Sun, Jian (6, 13); Wang, Hua (9, 35); Wang, Yinhai (14, 26); Wang, Yunpeng (6, 25); Yang, Hai (24, 131); Yin, Yafeng (10, 38); Yu, Bin (6, 11); Zhang, Guohui (6, 15); Zhang, H. M. (9, 24); Zhang, Junyi (3, 6); Zhang, Xiaoning (12, 28) |
| #5 (n=21) | Axhausen, Kay W. (21, 34); Bhat, Chandra R. (12, 56); Bliemer, Michiel C. J. (12, 28); Chiou, Yu-Chiun (1, 5); Corman, Francesco (3, 3); Golob, TF (2, 3); Hensher, David A. (18, 119); Jara-Diaz, Sergio (4, 8); Jou, Rong-Chang (5, 6); Kitamura, R (7, 7); Koppelman, FS (2, 6); Lee, Der-Horng (5, 11); Li, Zheng (5, 34); Mahmassani, Hani S. (4, 9); Menendez, Monica (5, 9); Mokhtarian, Patricia L. (9, 14); Pendyala, Ram M. (6, 30); Rose, John M. (11, 72); Tirachini, Alejandro (9, 19); Viti, Francesco (5, 5); Zhou, Xuesong (12, 21) |
| #6 (n=19) | Ban, Xuegang (Jeff) (8, 18); Cantillo, Victor (8, 26); Chen, Xiqun (Michael) (8, 18); Dell'olio, Luigi (3, 31); El-Geneidy, Ahmed (6, 21); Fu, Liping (5, 15); Holguin-Veras, Jose (7, 11); Ibeas, Angel (5, 33); Levinson, David (5, 13); Li, Li (13, 35); Liu, Henry X. (9, 20); Lord, Dominique (8, 17); Ma, Wanjing (7, 13); Manaugh, Kevin (3, 15); Miranda-Moreno, Luis F. (5, 20); Ozbay, Kaan (2, 4); Ran, Bin (9, 15); Zhang, Lei (6, 16); Zhang, Yunlong (2, 7) |
| #7 (n=18) | Chatterjee, Kiron (2, 6); Daly, Andrew (5, 26); Hess, Stephane (12, 42); Ison, Stephen (8, 18); Liu, Yang (8, 11); Lucas, Karen (6, 10); Lyons, Glenn (4, 8); Marsden, Greg (5, 8); Merkert, Rico (7, 13); Monios, Jason (3, 7); Mulley, Corinne (8, 31); Preston, John (2, 2); Quddus, Mohammed (13, 30); Rye, Tom (4, 10); Ryley, Tim (4, 6); Scheiner, Joachim (1, 1); Wang, Chao (6, 15); Wardman, Mark (5, 7) |
| #8 (n=18) | Crundall, David (1, 1); Gkritza, Konstantina (4, 9); Hansen, Mark (1, 2); Huang, Helai (12, 52); Karlaftis, Matthew G. (5, 13); Li, Ye (10, 34); Li, Zhibin (9, 52); Liu, Pan (11, 63); Mannering, Fred (9, 17); Peeta, Srinivas (4, 4); Schwebel, David C. (2, 2); Ukkusuri, Satish V. (4, 4); Vlahogianni, Eleni I. (4, 14); Wang, Hao (12, 35); Wang, Wei (14, 101); Xu, Chengcheng (8, 61); Zhang, Wei (9, 13); Zhang, Yu (4, 8) |
| #9 (n=17) | Bekhor, Shlomo (7, 14); Cassidy, Michael J. (3, 9); Daganzo, CF (6, 13); Farah, Haneen (7, 23); Geroliminis, Nikolas (13, 20); Hall, RW (1, 2); Hoogendoorn, Serge P. (15, 55); Kaplan, Sigal (7, 38); Knoop, Victor L. (9, 27); Leclercq, Ludovic (3, 12); Li, Xiaopeng (10, 26); Ouyang, Yanfeng (8, 19); Prato, Carlo G. (8, 43); Shiftan, Yoram (6, 12); Taubman-Ben-Ari, Orit (4, 12); Toledo, Tomer (11, 23); Van Arem, Bart (7, 26) |
| #10 (n=15) | Brown, Julie (1, 7); Cats, Oded (10, 27); Delhomme, Patricia (2, 9); Eluru, Naveen (13, 42); Haworth, Narelle (8, 18); Ivers, Rebecca (2, 9); King, Mark (17, 32); Oviedo-Trespalacios, Oscar (6, 18); Rakotonirainy, Andry (7, 18); Shinar, David (2, 2); Susilo, Yusak O. (5, 11); Tay, Richard (2, 3); Washington, Simon (11, 30); Watson, Barry (7, 8); Zheng, Zuduo (4, 16) |
| #11 (n=13) | Bedard, Michel (3, 5); Ceder, Avishai (Avi) (4, 4); Currie, Graham (10, 48); Delbosc, Alexa (5, 31); Eby, David W. (2, 28); Gabler, Hampton C. (1, 1); Koppel, Sjaan (5, 24); Lenne, Michael G. (5, 6); Molnar, Lisa J. (3, 30); Newstead, Stuart (1, 4); Sarvi, Majid (7, 18); Stephens, Amanda N. (4, 19); Sullman, Mark J. M. (1, 10) |
| #12 (n=11) | Banister, David (3, 9); Button, K (2, 3); De Vos, Jonas (7, 28); Dijst, Martin (3, 14); Ettema, Dick (10, 62); Givoni, Moshe (3, 9); Nijkamp, P (3, 25); Rietveld, Piet (6, 33); Schwanen, Tim (9, 23); Verhoef, Erik T. (1, 1); Witlox, Frank (7, 30) |
| #13 (n=10) | Chen, Anthony (11, 27); Choi, Keechoo (14, 35); Lam, William H. K. (23, 112); Li, Zhi-Chun (10, 48); Lo, Hong K. (18, 51); Loo, Becky P. Y. (7, 12); Sumalee, Agachai (7, 28); Sze, N. N. (11, 32); Wang, Donggen (3, 5); Wong, S. C. (21, 138) |
| #14 (n=9) | Bell, Michael G. H. (16, 41); Friman, Margareta (3, 39); Fujii, Satoshi (9, 32); Garling, Tommy (4, 42); Graham, Daniel J. (4, 7); Morikawa, Takayuki (3, 14); Noland, Robert B. (4, 9); Schmoecker, Jan-Dirk (7, 20); Yamamoto, Toshiyuki (3, 13) |
| #15 (n=8) | Gao, Ziyou (23, 97); Liu, Ronghui (15, 31); Papageorgiou, Markos (4, 27); Papamichail, Ioannis (1, 24); Sun, Huijun (8, 45); Wang, David Z. W. (9, 30); Wu, Jianjun (12, 53); Yang, Lixing (7, 35) |
| #16 (n=7) | Laporte, Gilbert (1, 2); Liu, Zhiyuan (17, 66); Meng, Qiang (14, 117); Qu, Xiaobo (9, 36); Wang, Shuaian (12, 86); Weng, Jinxian (7, 35); Yan, Xuedong (16, 37) |
| #17 (n=6) | Abdel-Aty, Mohamed (17, 117); El-Basyouny, Karim (1, 10); Lee, Jaeyoung (8, 66); Sayed, Tarek (3, 12); Wang, Xuesong (10, 40); Yu, Rongjie (7, 37) |
| #18 (n=6) | Cicchino, Jessica B. (1, 6); Ferguson, SA (3, 35); Lund, AK (3, 25); Mccartt, Anne T. (3, 14); Preusser, DF (3, 44); Williams, AF (5, 74) |
| #19 (n=4) | Brijs, Kris (3, 63); Brijs, Tom (3, 99); Daniels, Stijn (6, 69); Wets, Geert (7, 97); |
| #20 (n=3) | Gatta, Valerio (2, 27); Macharis, Cathy (2, 4); Marcucci, Edoardo (3, 28) |
| #21 (n=2) | Malighetti, Paolo (1, 23); Redondi, Renato (2, 24) |

## Appendix E – Most cited transportation papers, based on global citation count

| Title | Authors (year) | Journal | Citation count |
|---|---|---|---|
| Travel demand and the 3Ds: Density, diversity, and design | Cervero and Kockelman (1997) | Transportation Research Part D | 1503 |
| The cell transmission model - a dynamic representation of highway traffic consistent with the hydrodynamic theory | Daganzo (1994a) | Transportation Research Part B | 1423 |
| The cell transmission model .2. Network traffic | Daganzo (1995a) | Transportation Research Part B | 1046 |
| The Mixed Logit model: The state of practice | Hensher and Greene (2003) | Transportation | 903 |
| A behavioral car-following model for computer-simulation | Gipps (1981) | Transportation Research Part B | 876 |
| The sustainable mobility paradigm | Banister (2008) | Transport Policy | 801 |
| The statistical analysis of crash-frequency data: A review and assessment of methodological alternatives | Lord and Mannering (2010) | Transportation Research Part A | 767 |
| Making cycling irresistible: Lessons from the Netherlands, Denmark and Germany | Pucher and Buehler (2008) | Transport Reviews | 703 |
| A latent class model for discrete choice analysis: Contrasts with mixed logit | Greene and Hensher (2003) | Transportation Research Part B | 693 |
| Network Design and Transportation-Planning - Models and Algorithms | Magnanti and Wong (1984) | Transportation Science | 679 |
| Self-organized pedestrian crowd dynamics: Experiments, simulations, and design solutions | Helbing et al. (2005) | Transportation Science | 676 |
| Preparing a nation for autonomous vehicles: opportunities, barriers and policy recommendations | Fagnant and Kockelman (2015) | Transportation Research Part A | 666 |
| An adaptive large neighborhood search heuristic for the pickup and delivery problem with time windows | Ropke and Pisinger (2006) | Transportation Science | 661 |
| Existence of urban-scale macroscopic fundamental diagrams: Some experimental findings | Geroliminis and Daganzo (2008) | Transportation Research Part B | 588 |
| Revenue management: Research overview and prospects | McGill and Van Ryzin (1999) | Transportation Science | 569 |
| Car use: lust and must. Instrumental, symbolic and affective motives for car use | Steg (2005) | Transportation Research Part A | 564 |
| A continuum theory for the flow of pedestrians | Hughes (2002) | Transportation Research Part B | 563 |
| A micro-analysis of land use and travel in five neighborhoods in the San Francisco Bay Area | Kitamura et al. (1997) | Transportation | 551 |
| Correlation or causality between the built environment and travel behavior? Evidence from Northern California | Handy et al. (2005) | Transportation Research Part D | 529 |
| Vehicle routing problem with time windows, part 1: Route construction and local search algorithms | Braysy and Gendreau (2005) | Transportation Science | 524 |
| Driving speed and the risk of road crashes: A review | Aarts and van Schagen (2006) | Accident Analysis and Prevention | 519 |
| Examining the Impacts of Residential Self-Selection on Travel Behaviour: A Focus on Empirical Findings | Cao et al. (2009b) | Transport Reviews | 511 |
| Comparison of parametric and nonparametric models for traffic flow forecasting | Smith et al. (2002) | Transportation Research Part B | 501 |
| A simplified theory of kinematic waves in highway traffic .1. General theory | Newell (1993) | Transportation Research Part B | 497 |
| Transport and climate change: a review | Chapman (2007) | Journal of Transport Geography | 492 |
| Commuting by Bicycle: An Overview of the Literature | Heinen et al. (2010) | Transport Reviews | 490 |
| Requiem for second-order fluid approximations of traffic flow | Daganzo (1995c) | Transportation Research Part B | 487 |
| Traffic Equilibrium and Variational-Inequalities | Dafermos (1980) | Transportation Science | 487 |
| A tabu search heuristic for the vehicle routing problem with soft time windows | Taillard et al. (1997) | Transportation Science | 480 |
| Long short-term memory neural network for traffic speed prediction using remote microwave sensor data | Ma et al. (2015) | Transportation Research Part C | 475 |

## Appendix F – Top cited papers of each 'Transportation' journal

| Journal title | Article | Cit. | Journal title | Article | Cit. | Journal title | Article | Cit. |
|---|---|---|---|---|---|---|---|---|
| Accid. Anal. Prev. (n=6,410) | Aarts and van Schagen (2006) | 514 | J. Air Transp. Manag. (n=1,559) | Liou et al. (2007) | 254 | Eur. Transp. Res. Rev. (n=345) | Kumar and Vanajakshi (2015) | 107 |
| | Zohar (2010) | 453 | | O'Connell and Williams (2005) | 178 | | Agostinacchio et al. (2014) | 76 |
| | Jonah (1997) | 430 | | Park et al. (2004) | 178 | | Schnebele et al. (2015) | 54 |
| Transp. Res. Pt. A-Policy Pract. (n=3,715) | Lord and Mannering (2010) | 764 | Transp. J. (n=1,432) | Stank et al. (2001) | 160 | Int. J. Shipp. Transp. Logist. (n=344) | Lam and Gu (2013) | 57 |
| | Fagnant and Kockelman (2015) | 648 | | Cruijssen et al. (2007) | 143 | | Wong (2009) | 48 |
| | Steg (2005) | 559 | | Maloni and Carter (2006) | 106 | | Bichou (2011) | 45 |
| Transp. Res. Pt. B-Methodol. (n=2,884) | Daganzo (1994a) | 1416 | J. Transp. Econ. Policy (n=1,364) | Goodwin (1992) | 316 | Eur. J. Transport. Infrastruct. Res. (n=333) | Chorus (2010) | 128 |
| | Daganzo (1995a) | 1043 | | Louviere (1988) | 254 | | Jorge and Correia (2013) | 116 |
| | Gipps (1981) | 873 | | Oum et al. (1992) | 249 | | van Wee and Geurs (2011) | 96 |
| Transport. Res. Part D-Transport. Environ. (n=2,318) | Cervero and Kockelman (1997) | 1495 | Transp. Rev. (n=1,358) | Pucher and Buehler (2008) | 701 | Int. J. Transp. Econ. (n=323) | De Langen and Pallis (2006) | 48 |
| | Handy et al. (2005) | 525 | | Cao et al. (2009b) | 508 | | Ferrari et al. (2010) | 35 |
| | Kempton and Letendre (1997) | 398 | | Heinen et al. (2010) | 485 | | Gonzalez-Feliu et al. (2013) | 34 |
| J. Transp. Geogr. (n=2,087) | Chapman (2007) | 486 | J. Transp. Health (n=1,008) | Goodman et al. (2014) | 70 | Marit. Econ. Logist. (n=318) | Alvarez (2009) | 101 |
| | Coffin (2007) | 456 | | Christiansen et al. (2016) | 67 | | Veenstra et al. (2012) | 88 |
| | Geels (2012) | 316 | | Griffin and Jiao (2015) | 51 | | Roso and Lumsden (2010) | 66 |
| J. Saf. Res. (n=1,973) | Williams (2003) | 395 | Marit. Policy Manag. (n=634) | Verhoeven (2010) | 139 | J. Transp. Land Use (n=281) | Cervero (2013) | 112 |
| | Dedobbeleer and Béland (1991) | 347 | | Rodrigue and Notteboom (2009) | 138 | | Daniels and Mulley (2013) | 88 |
| | Cooper and Phillips (2004) | 344 | | Eide et al. (2011) | 98 | | Nss (2012) | 73 |
| Transportation (n=1,911) | Hensher and Greene (2003) | 902 | Int. J. Sustain. Transp. (n=614) | Cervero et al. (2009) | 328 | Travel Behav. Soc. (n=249) | Yue et al. (2014) | 93 |
| | Kitamura et al. (1997) | 550 | | Shaheen and Cohen (2013) | 215 | | Popovich et al. (2014) | 64 |
| | Hensher (1994) | 293 | | Nair et al. (2013) | 160 | | Alemi et al. (2018) | 58 |
| Transp. Res. Pt. E-Logist. Transp. Rev. (n=1,895) | Bates et al. (2001) | 433 | Res. Transp. Econ. (n=590) | Pradhan and Bagchi (2013) | 86 | Res. Transp. Bus. Manag. (n=229) | Ricci (2015) | 114 |
| | Meixell and Gargeya (2005) | 373 | | Meyer et al. (2017) | 75 | | Hagman et al. (2016) | 65 |
| | Sheu (2007) | 349 | | Hidalgo and Gutierrez (2013) | 56 | | Schliwa et al. (2015) | 59 |
| Transp. Res. Pt. F-Traffic Psychol. Behav. (n=1,789) | Engström et al. (2005) | 346 | Mobilities (n=512) | Laurier et al. (2008) | 182 | J. Publ. Transp. (n=209) | Wang et al. (2011) | 68 |
| | Kyriakidis et al. (2015) | 296 | | Jensen (2009) | 153 | | Collins et al. (2013) | 60 |
| | Victor et al. (2005) | 228 | | Harvey and Knox (2012) | 127 | | Truong and Somenahalli (2011) | 52 |
| Traffic Inj. Prev. (n=1,788) | Molnar and Eby (2008) | 126 | J. Intell. Transport. Syst. (n=463) | Balcik et al. (2008) | 229 | Transportmetrica B-Transp. Dyn. (n=195) | Shepherd (2014) | 95 |
| | Shechtman et al. (2009) | 122 | | Trepanier et al. (2007) | 152 | | Watling and Cantarella (2013) | 57 |
| | McCartt et al. (2009) | 115 | | Milakis et al. (2017) | 150 | | Taylor (2013) | 47 |
| Transp. Sci. (n=1,637) | Magnanti and Wong (1984) | 674 | Transportmetrica A (n=424) | Daziano and Bolduc (2013) | 96 | Anal. Methods Accid. Res. (n=99) | Mannering et al. (2016) | 318 |
| | Helbing et al. (2005) | 670 | | Mori et al. (2015) | 90 | | Behnood and Mannering (2015) | 81 |
| | Ropke and Pisinger (2006) | 655 | | Bordagaray et al. (2014) | 67 | | Barua et al. (2016) | 79 |
| Transp. Policy (n=1,600) | Banister (2008) | 793 | Transp. Lett. (n=365) | Hood et al. (2011) | 156 | Econ. Transp. (n=96) | Inci (2015) | 66 |
| | Anable (2005) | 461 | | Mokhtarian (2009) | 85 | | Fosgerau (2015) | 38 |
| | Beirao and Cabral (2007) | 438 | | Auld et al. (2009) | 72 | | Small (2015) | 36 |
| Transp. Res. Pt. C-Emerg. Technol. (n=2692) | Smith et al. (2002) | 501 | J. Transp. Saf. Secur. (n=317) | Yasmin et al. (2014) | 29 | J. Adv. Transp. (n=159) | Chen et al. (1999) | 159 |
| | Ma et al. (2015) | 475 | | Shi et al. (2014) | 21 | | Imai et al. (1997) | 148 |
| | Quddus et al. (2007) | 463 | | Cosma et al. (2016) | 19 | | Bejan (1996) | 141 |

## Appendix G – Most cited reviews of transportation literature, based on global citation count

| Title | Authors (year) | Journal | Citation count |
|---|---|---|---|
| The statistical analysis of crash-frequency data: A review and assessment of methodological alternatives | Lord and Mannering (2010) | Transportation Research Part A | 767 |
| Driving speed and the risk of road crashes: A review | Aarts and van Schagen (2006) | Accident Analysis and Prevention | 519 |
| Transport and climate change: a review | Chapman (2007) | Journal of Transport Geography | 492 |
| From roadkill to road ecology: A review of the ecological effects of roads | Coffin (2007) | Journal of Transport Geography | 460 |
| Models and algorithms for road network design: a review and some new developments | Yang and Bell (1998) | Transport Reviews | 446 |
| Sensation seeking and risky driving: A review and synthesis of the literature | Jonah (1997) | Accident Analysis and Prevention | 430 |
| The statistical analysis of highway crash-injury severities: A review and assessment of methodological alternatives | Savolainen et al. (2011) | Accident Analysis and Prevention | 412 |
| Smart card data use in public transit: A literature review | Pelletier et al. (2011) | Transportation Research Part C | 403 |
| Global supply chain design: A literature review and critique | Meixell and Gargeya (2005) | Transportation Research Part E | 374 |
| Elasticities of road traffic and fuel consumption with respect to price and income: A review | Goodwin et al. (2004) | Transport Reviews | 367 |
| A meta-analysis of the effects of cell phones on driver performance | Caird et al. (2008) | Accident Analysis and Prevention | 323 |
| Advances in consumer electric vehicle adoption research: A review and research agenda | Rezvani et al. (2015) | Transportation Research Part E | 318 |
| A review of new demand elasticities with special reference to short and long-run effects of price changes | Goodwin (1992) | Journal of Transport Economics and Policy | 316 |
| Bikeshare: A Review of Recent Literature | Fishman (2016) | Transport Reviews | 273 |
| Conjoint-analysis modeling of stated preferences - a review of theory, methods, recent developments and external validity | Louviere (1988) | Journal of Transport Economics and Policy | 254 |
| Development and impact of the modern high-speed train: A review | Givoni (2006) | Transport Reviews | 253 |
| Constructing Efficient Stated Choice Experimental Designs | Rose and Bliemer (2009) | Transport Reviews | 252 |
| Models of driving behavior - A review of their evolution | Ranney (1994) | Accident Analysis and Prevention | 251 |
| Is a new applied transportation research field emerging? - A review of intermodal rail-truck freight transport literature | Bontekoning et al. (2004) | Transportation Research Part A | 239 |
| Bicycle helmet efficacy: a meta-analysis | Attewell et al. (2001) | Accident Analysis and Prevention | 234 |
| The Driver Behaviour Questionnaire as a predictor of accidents: A meta-analysis | de Winter and Dodou (2010) | Journal of Safety Research | 228 |
| Thirty Years of Inventory Routing | Coelho et al. (2014) | Transportation Science | 224 |
| Travel time variability: a review of theoretical and empirical issues | Noland and Polak (2002) | Transport Reviews | 219 |
| Have young workers more injuries than older ones? An international literature review | Salminen (2004) | Journal of Safety Research | 215 |
| A review of neural networks applied to transport. | Dougherty (1995) | Transportation Research Part C | 212 |
| A review of risk factors and patterns of motorcycle injuries | Lin and Kraus (2009) | Accident Analysis and Prevention | 209 |
| Effects of adaptive cruise control and highly automated driving on workload and situation awareness: A review of the empirical evidence | de Winter et al. (2014) | Transportation Research Part F | 207 |
| Road traffic demand elasticity estimates: A review | Graham and Glaister (2004) | Transport Reviews | 207 |
| Planning, operation, and control of bus transport systems: A literature review | Ibarra-Rojas et al. (2015) | Transportation Research Part B | 202 |
| Telecommunications and travel relationships - a review | Salomon (1986) | Transportation Research Part A | 200 |

**Appendix H – The extent of activities in various streams of transportation research clusters (1970-2020) as reflected in instances of co-citations. Icons • and •• respectively represent moderate and intense activities (of clusters during each year). The colour-coding is consistent with that of Figures 4, 5 and 6.**

| ID | Cluster label | Broader affiliation | 1970 | 1971 | 1972 | 1973 | 1974 | 1975 | 1976 | 1977 | 1978 | 1979 | 1980 | 1981 | 1982 | 1983 | 1984 | 1985 | 1986 | 1987 | 1988 | 1989 | 1990 | 1991 | 1992 | 1993 | 1994 | 1995 | 1996 | 1997 | 1998 | 1999 | 2000 | 2001 | 2002 | 2003 | 2004 | 2005 | 2006 | 2007 | 2008 | 2009 | 2010 | 2011 | 2012 | 2013 | 2014 | 2015 | 2016 | 2017 | 2018 | 2019 | 2020 |
|---|---|---|---|---|---|---|---|---|---|---|---|---|---|---|---|---|---|---|---|---|---|---|---|---|---|---|---|---|---|---|---|---|---|---|---|---|---|---|---|---|---|---|---|---|---|---|---|---|---|---|---|---|---|
| 0 | congested network | #1 | | | | | | | | | | •• | •• | •• | •• | •• | •• | •• | •• | •• | •• | • | •• | • | •• | •• | •• | •• | •• | •• | •• | •• | • | •• | •• | | •• | •• | • | | | | • | • | • | • | • | • | | | | • | |
| 1 | driving anger | #4 | | | | | | | | | | | | | | | | | | | | •• | • | | | • | • | • | •• | • | •• | • | • | • | •• | •• | •• | •• | •• | •• | •• | • | • | •• | •• | •• | •• | • | •• | • | • | | |
| 2 | prospect theory | #2,3 | | | | | | | | | | | | | | | • | | | • | • | • | | | • | • | • | | | • | • | •• | •• | •• | | | •• | •• | •• | | • | •• | •• | • | •• | • | •• | • | | | • | | |
| 3 | integrating theme | #3 | | | | | | | | • | •• | • | • | • | •• | •• | • | • | | • | | • | | | • | | | | | | | | | • | • | | | | • | • | | | | | | | • | | | | | | |
| 4 | random parameter | #4 | | | | | | | | | | | | | | | | | | | | | | | | | | | | | | | | | | | • | | | | | •• | | •• | • | •• | •• | • | •• | •• | •• | •• | •• |
| 5 | travel behaviour | #3 | | | | | | | | | | | | | | | | | | | | | | | | | | | | | | | | | | | | | | • | | • | | • | • | • | •• | •• | •• | •• | •• | •• | • |
| 6 | delivery problem | #1,2 | | | | | | | | | | | • | • | | | | •• | | | | | | • | •• | •• | | • | | | • | | • | | | | •• | • | • | • | | | | | | | | | • | | | | |
| 7 | motor vehicle | #4 | | • | | | | • | | • | • | | | | | • | • | • | • | • | •• | •• | •• | • | • | •• | • | •• | • | | | | | | | | • | • | | | | | | | | | | | | | | | |
| 8 | technical efficiency | #2 | | | | | | | | | | • | • | | | | | | | • | | | •• | | | • | | | | • | | | | | • | | | | • | | • | •• | | •• | | | | | | | | | |
| 9 | public transport | #3 | | | | | | | | | | | | | | | | | | | | | | | | | | | | | | | | | | • | | | | • | • | • | • | • | •• | • | •• | •• | •• | •• | •• | •• | • |
| 10 | automated vehicle | #1 | | | | | | | | | | | | | | | | | | | | | | | | | | | | | | | | | | | | | | | | | | | | | | • | • | • | • | • | •• |
| 11 | liner shipping | #2 | | | | | | | | | | | | | | | | | | | | | | | | | | | | | | | | | | | | | | | | | | | • | • | • | •• | • | • | | • | • |
| 12 | deviation concept | #4 | | | | •• | | • | •• | | | | | | | | •• | | | | | | | | | | | | | | | | | | | | | | | | | | | | | | | | | | | | |
| 13 | driving avoidance | #4 | | | | | | | | | | | | | | | | | • | • | | • | • | | | • | | | • | • | | | | | •• | •• | | | • | | | | | | | • | | | | | | | |
| 14 | safety behaviour | #4 | | | | | | | | | | | | | | | | | | | | | | | | | | | | | | | | | • | | • | | | • | | | •• | | •• | | | | | | | | |
| 15 | mobile phone | #4 | | | | | | | | | | | | | | | | | | | | | | | | | | | | | | | | | | | | • | • | • | •• | •• | • | • | • | • | • | • | • | • | • | •• | • |
| 16 | autonomous vehicle | #3,2 | | | | | | | | | | | | | | | | | | | | | | | | | | | | | | | | | | | | | | | | | | | • | | | • | • | • | •• | •• | • |
| 17 | social exclusion | #3 | | | | | | | | | | | | | | | | | | | | | | | | | | | | | | | | | | | | | | | | | • | • | | | | • | • | • | •• | •• | • |
| 18 | macroscopic fundamental diagram | #1 | | | | | | | | | | | | | | | | | | | | | | | | | | | | | | | | | • | •• | | | | | | | | • | • | • | | •• | •• | • | •• | • | •• |
| 19 | bus bunching | #1,3 | | | | | | | | | | | | | | | | | | | | | | | | | | | | | | | | | | | | | | | | | | | | | | | • | • | | | |
| 21 | electric vehicle | #2,1 | | | | | | | | | | | | | | | | | | | | | | | | | | | | | | | | | | | | | | | | | | | | | | • | • | •• | •• | •• | • |
| 22 | mass transit | #2 | | • | | • | •• | • | • | | | • | | | | •• | | • | • | | • | • | | | | | | • | | | | | | | | | | | | | | | | | | | | | | | | | |
| 23 | shared mobility | #1,3 | | | | | | | | | | | | | | | | | | | | | | | | | | | | | | | | | | | | | | | | | | | | | | | | • | | •• | • |

# Appendix I – Influential outsiders of transportation literature

| Division | Author(s) (year) | Title | Journal | Cit. |
|---|---|---|---|---|
| Network modelling and traffic flow | Richards (1956) | Shock Waves on the Highway | Operations Research | 455 |
| | Lighthill and Whitham (1955) | On kinematic waves II. A theory of traffic flow on long crowded roads | Proceedings of the Royal Society of London. Series A. Mathematical and Physical Sciences | 330 |
| | Wardrop (1952) | Road paper. some theoretical aspects of road traffic research | Proceedings of the institution of civil engineers | 299 |
| | Treiber et al. (2000) | Congested traffic states in empirical observations and microscopic simulations | Physical Review E | 159 |
| | Arnott et al. (1993) | A structural model of peak-period congestion: A traffic bottleneck with elastic demand | The American Economic Review | 152 |
| | Van Arem et al. (2006) | The impact of cooperative adaptive cruise control on traffic-flow characteristics | IEEE Transactions on Intelligent Transportation Systems | 104 |
| | Agatz et al. (2012) | Optimization for dynamic ride-sharing: A review | European Journal of Operational Research | 89 |
| | Friesz et al. (1993) | A variational inequality formulation of the dynamic network user equilibrium problem | Operations Research | 89 |
| | Farahani et al. (2013) | A review of urban transportation network design problems | European Journal of Operational Research | 47 |
| Economics of transport | Vickrey (1969) | Congestion theory and transport investment | The American Economic Review | 420 |
| | Small (1982) | The scheduling of consumer activities: work trips | The American Economic Review | 303 |
| | Charnes et al. (1978) | Measuring the efficiency of decision making units | European Journal of Operational Research | 298 |
| | Kahneman (1979) | Prospect theory: An analysis of decisions under risk | Econometrica | 291 |
| | McFadden and Train (2000) | Mixed MNL models for discrete response | Journal of Applied Econometrics | 264 |
| | Banker et al. (1984) | Some models for estimating technical and scale inefficiencies in data envelopment analysis | Management Science | 128 |
| | Mohring (1972) | Optimization and scale economies in urban bus transportation | The American Economic Review | 114 |
| | Egbue and Long (2012) | Barriers to widespread adoption of electric vehicles: An analysis of consumer attitudes and perceptions | Energy Policy | 108 |
| | Hidrue et al. (2011) | Willingness to pay for electric vehicles and their attributes | Resource and energy economics | 100 |
| | Farrell (1957) | The measurement of productive efficiency | Journal of the Royal Statistical Society: Series A (General) | 91 |
| | Caves et al. (1984) | Economies of density versus economies of scale: why trunk and local service airline costs differ | The RAND Journal of Economics | 84 |
| | Williams (1977) | On the formation of travel demand models and economic evaluation measures of user benefit | Environment and Planning A | 80 |
| | Borenstein (1989) | Hubs and high fares: dominance and market power in the US airline industry | The RAND Journal of Economics | 78 |
| | Davis (1989) | Perceived Usefulness, Perceived Ease of Use, and User Acceptance of Information Technology | MIS Quarterly | 75 |
| | Small et al. (2005) | Uncovering the distribution of motorists' preferences for travel time and reliability | Econometrica | 59 |
| | DeSerpa (1971) | A theory of the economics of time | The Economic Journal | 31 |
| Logistics | Solomon (1987) | Algorithms for the vehicle routing and scheduling problems with time window constraints | Operations Research | 146 |
| | Christiansen et al. (2013) | Ship routing and scheduling in the new millennium | European Journal of Operational Research | 68 |
| | Bodin (1983) | Routing and scheduling of vehicles and crews, the state of the art | Computers & Operations Research | 54 |
| | Hakimi (1964) | Optimum locations of switching centers and the absolute centers and medians of a graph | Operations Research | 22 |
| Travel behaviour | Ewing and Cervero (2010) | Travel and the built environment: a meta-analysis | Journal of the American Planning Association | 511 |
| | Hansen (1959) | How accessibility shapes land use | Journal of the American Institute of Planners | 238 |
| | Sheller and Urry (2006) | The new mobilities paradigm | Environment and Planning A | 224 |
| | Pucher et al. (2010) | Infrastructure, programs, and policies to increase bicycling: an international review | Preventive Medicine | 151 |
| | Handy and Niemeier (1997) | Measuring accessibility: an exploration of issues and alternatives | Environment and Planning A | 131 |
| | Saelens et al. (2003) | Environmental correlates of walking and cycling: findings from the transportation, urban design, and planning literatures | Annals of behavioral medicine | 131 |
| | De Hartog et al. (2010) | Do the health benefits of cycling outweigh the risks? | Environmental Health Perspectives | 129 |
| | Bamberg and Schmidt (2003) | Incentives, morality, or habit? Predicting students' car use for university routes with the models of Ajzen, Schwartz, and Triandis | Environment and Behavior | 119 |
| | Cresswell (2010) | Towards a politics of mobility | Environment and Planning D: Society and Space | 106 |
| | Oja et al. (2011) | Health benefits of cycling: a systematic review | Scandinavian Journal of Medicine & Science in Sports | 102 |
| | Saelens and Handy (2008) | Built environment correlates of walking: a review | Medicine and science in sports and exercise | 92 |
| | Garrard et al. (2008) | Promoting transportation cycling for women: the role of bicycle infrastructure | Preventive Medicine | 77 |
| | Woodcock et al. (2009) | Public health benefits of strategies to reduce greenhouse-gas emissions: urban land transport | The Lancet | 66 |
| | Pucher and Dijkstra (2003) | Promoting safe walking and cycling to improve public health: lessons from the Netherlands and Germany | American Journal of Public Health | 53 |
| | Mueller et al. (2015) | Health impact assessment of active transportation: a systematic review | Preventive Medicine | 50 |
| | Parasuraman et al. (1988) | Servqual: A multiple-item scale for measuring consumer perc | Journal of Retailing | 59 |
| Road safety | Ajzen (1991) | The theory of planned behavior | Organizational Behavior and Human Decision Processes | 721 |
| | Reason et al. (1990) | Errors and violations on the roads: a real distinction? | Ergonomics | 345 |
| | Hu and Bentler (1999) | Cutoff criteria for fit indexes in covariance structure analysis: Conventional criteria versus new alternatives | Structural equation modeling: a multidisciplinary journal | 255 |
| | Fornell and Larcker (1981) | Evaluating structural equation models with unobservable variables and measurement error | Journal of Marketing Research | 223 |
| | Jacobsen (2015) | Safety in numbers: more walkers and bicyclists, safer walking and bicycling | Injury Prevention | 179 |
| | Anderson and Gerbing (1988) | Structural equation modeling in practice: A review and recommended two-step approach | Psychological bulletin | 178 |
| | Podsakoff et al. (2003) | Common method biases in behavioral research: a critical review of the literature and recommended remedies | Journal of Applied Psychology | 172 |
| | Spiegelhalter et al. (2002) | Bayesian measures of model complexity and fit | Journal of the Royal Statistical Society: Series B (Statistical Methodology) | 161 |
| | Parker et al. (1995) | Driving errors, driving violations and accident involvement | Ergonomics | 158 |
| | Armitage and Conner (2001) | Efficacy of the theory of planned behaviour: A meta-analytic review | British journal of social psychology | 149 |
| | Elander et al. (1993) | Behavioral correlates of individual differences in road-traffic crash risk | Psychological bulletin | 148 |
| | Ulleberg and Rundmo (2003) | Personality, attitudes and risk perception as predictors of risky driving behaviour among young drivers | Safety Science | 139 |
| | Klauer et al. (2014) | Distracted driving and risk of road crashes among novice and experienced drivers | New England Journal of Medicine | 122 |
| | Lawton et al. (1997) | The role of affect in predicting social behaviors: The case of road traffic violations | Journal of Applied Psychology | 119 |
| | Deffenbacher et al. (1994) | Development of a driving anger scale | Psychological Reports | 115 |
| | Redelmeier and Tibshirani (1997) | Association between cellular-telephone calls and motor vehicle collisions | New England Journal of Medicine | 101 |
| | Anstey et al. (2005) | Cognitive, sensory and physical factors enabling driving safety in older adults | Clinical Psychology Review | 98 |
| | Baron and Kenny (1986) | The moderator–mediator variable distinction in social psychological research: Conceptual, strategic, and statistical considerations | Journal of Personality and Social Psychology | 97 |
| | Zohar (1980) | Safety climate in industrial organizations: theoretical and applied implications | Journal of Applied Psychology | 96 |
| | Dingus et al. (2016) | Driver crash risk factors and prevalence evaluation using naturalistic driving data | Proceedings of the National Academy of Sciences | 86 |
| | Endsley (1995) | Toward a theory of situation awareness in dynamic systems | Human Factors | 78 |
| | Reynolds et al. (2009) | The impact of transportation infrastructure on bicycling injuries and crashes: a review of the literature | Environmental health | 74 |
| | Chen et al. (2000) | Carrying passengers as a risk factor for crashes fatal to 16-and 17-year-old drivers | JAMA | 54 |
| | Parker et al. (1992) | Intention to commit driving violations: An application of the theory of planned behavior | Journal of Applied Psychology | 47 |
| | Baker et al. (1974) | The injury severity score: a method for describing patients with multiple injuries and evaluating emergency care | Journal of Trauma and Acute Care Surgery | 41 |